\documentclass[longauth]{aa}
\usepackage{graphicx}
\usepackage{txfonts}
\usepackage{xcolor}
\usepackage{hyperref}
\hypersetup{colorlinks, linkcolor=blue, citecolor=blue, urlcolor=blue}

\newcommand{\pivec}{\mbox{\boldmath $\pi$}}
\newcommand{\muvec}{\mbox{\boldmath $\mu$}}

\newcommand{\te}{t_{\rm E}}
\newcommand{\thetae}{\theta_{\rm E}}

\newcommand{\pie}{\pi_{\rm E}}
\newcommand{\pien}{\pi_{{\rm E},N}}
\newcommand{\piee}{\pi_{{\rm E},E}}
\newcommand{\dl}{D_{\rm L}}
\newcommand{\ds}{D_{\rm S}}
\newcommand{\hjd}{{\rm HJD}}

\definecolor{brown}{rgb}{0.59, 0.29, 0.0}
\definecolor{darkgreen}{rgb}{0.0, 0.42, 0.24}
\definecolor{darkblue}{rgb}{0.01, 0.31, 0.59}
\definecolor{darkblue}{rgb}{0.0, 0.25, 0.42}
\definecolor{blue}{rgb}{0.0,0.0,1.0}
\definecolor{green}{rgb}{0.0,1.0,0.0}

\def\eqalign#1{\null\,\vcenter{\openup\jot
        \ialign{\strut\hfil$\displaystyle{##}$&$
        \displaystyle{{}##}$\hfil \crcr#1\crcr}}\,}

\begin{document}

\title{
Analyses of anomalous lensing events detected from the UKIRT microlensing survey
}
\titlerunning{UKIRT anomalous lensing events}

\author{
     Cheongho~Han\inst{\ref{inst01}} 
\and Weicheng~Zang\inst{\ref{inst07},\ref{inst08}}     
\and Andrzej~Udalski\inst{\ref{inst04}} 
\and Chung-Uk~Lee\inst{\ref{inst05}\thanks{Corresponding author.}} 
\and Ian~A.~Bond\inst{\ref{inst06}}
\and Yongxin Wen \inst{\ref{inst02},\ref{inst03}}
\and Bo Ma\inst{\ref{inst02},\ref{inst03}}
\\
(Leading authors)
\\
     Michael~D.~Albrow\inst{\ref{inst09}}   
\and Sun-Ju~Chung\inst{\ref{inst05}}      
\and Andrew~Gould\inst{\ref{inst10}}      
\and Kyu-Ha~Hwang\inst{\ref{inst05}} 
\and Youn~Kil~Jung\inst{\ref{inst05}} 
\and Yoon-Hyun~Ryu\inst{\ref{inst05}} 
\and Yossi~Shvartzvald\inst{\ref{inst11}}   
\and In-Gu~Shin\inst{\ref{inst07}} 
\and Hongjing Yang\inst{\ref{inst08}}
\and Jennifer~C.~Yee\inst{\ref{inst07}}   
\and Doeon~Kim\inst{\ref{inst01}}
\and Dong-Jin~Kim\inst{\ref{inst05}} 
\and Sang-Mok Cha\inst{\ref{inst05},\ref{inst31}}
\and Seung-Lee Kim\inst{\ref{inst05}}
\and Dong-Joo Lee\inst{\ref{inst05}}
\and Yongseok Lee\inst{\ref{inst05},\ref{inst31}}
\and Byeong-Gon~Park\inst{\ref{inst05}} 
\and Richard~W.~Pogge\inst{\ref{inst10}}
\\
(The KMTNet Collaboration)
\\
     Przemek~Mr{\'o}z\inst{\ref{inst04}} 
\and Micha{\l}~K.~Szyma{\'n}ski\inst{\ref{inst04}}
\and Jan~Skowron\inst{\ref{inst04}}
\and Rados{\l}aw~Poleski\inst{\ref{inst04}} 
\and Igor~Soszy{\'n}ski\inst{\ref{inst04}}
\and Pawe{\l}~Pietrukowicz\inst{\ref{inst04}}
\and Szymon~Koz{\l}owski\inst{\ref{inst04}} 
\and Krzysztof~A.~Rybicki\inst{\ref{inst04},\ref{inst11}}
\and Patryk~Iwanek\inst{\ref{inst04}}
\and Krzysztof~Ulaczyk\inst{\ref{inst12}}
\and Marcin~Wrona\inst{\ref{inst04},\ref{inst13}}
\and Mariusz~Gromadzki\inst{\ref{inst04}}          
\and Mateusz~J.~Mr{\'o}z\inst{\ref{inst04}} 
\\
(The OGLE Collaboration)
\\
     Fumio~Abe\inst{\ref{inst14}}
\and Ken~Bando\inst{\ref{inst19}}
\and David~P.~Bennett\inst{\ref{inst15},\ref{inst16}}
\and Aparna~Bhattacharya\inst{\ref{inst15},\ref{inst16}}
\and Akihiko~Fukui\inst{\ref{inst17},}\inst{\ref{inst18}}
\and Ryusei~Hamada\inst{\ref{inst19}}
\and Shunya~Hamada\inst{\ref{inst19}}
\and Naoto Hamasaki\inst{\ref{inst19}}
\and Yuki~Hirao\inst{\ref{inst25}}
\and Stela~Ishitani Silva\inst{\ref{inst15},\ref{inst20}}  
\and Naoki~Koshimoto\inst{\ref{inst19}}
\and Yutaka~Matsubara\inst{\ref{inst14}}
\and Shota~Miyazaki\inst{\ref{inst26}}
\and Yasushi~Muraki\inst{\ref{inst14}}
\and Tutumi Nagai\inst{\ref{inst19}}
\and Kansuke Nunota\inst{\ref{inst19}}
\and Greg~Olmschenk\inst{\ref{inst15}}
\and Cl{\'e}ment~Ranc\inst{\ref{inst22}}
\and Nicholas~J.~Rattenbury\inst{\ref{inst23}}
\and Yuki~Satoh\inst{\ref{inst19}}
\and Takahiro~Sumi\inst{\ref{inst19}}
\and Daisuke~Suzuki\inst{\ref{inst19}}
\and Sean K. Terry\inst{\ref{inst15}, \ref{inst16}}
\and Paul~J.~Tristram\inst{\ref{inst24}}
\and Aikaterini~Vandorou\inst{\ref{inst15},\ref{inst16}}
\and Hibiki~Yama\inst{\ref{inst19}}
\\
(The MOA Collaboration)
}

\institute{
      Department of Physics, Chungbuk National University, Cheongju 28644, Republic of Korea                                                          \label{inst01}  
\and  Center for Astrophysics $|$ Harvard \& Smithsonian 60 Garden St., Cambridge, MA 02138, USA                                                      \label{inst07}  
\and  Department of Astronomy, Tsinghua University, Beijing 100084, China                                                                             \label{inst08}  
\and  Astronomical Observatory, University of Warsaw, Al.~Ujazdowskie 4, 00-478 Warszawa, Poland                                                      \label{inst04}  
\and  Korea Astronomy and Space Science Institute, Daejon 34055, Republic of Korea                                                                    \label{inst05}  
\and  Institute of Natural and Mathematical Science, Massey University, Auckland 0745, New Zealand\label{inst3}                                       \label{inst06}  
\and  School of Physics and Astronomy, Sun Yat-sen University, Zhuhai 519082, People's Republic of China                                              \label{inst02}  
\and  CSST Science Center for the Guangdong-Hong Kong-Macau Great Bay Area, Sun Yat-sen University, Zhuhai 519082, China                              \label{inst03}  
\and  University of Canterbury, Department of Physics and Astronomy, Private Bag 4800, Christchurch 8020, New Zealand                                 \label{inst09}  
\and  Department of Astronomy, The Ohio State University, 140 W. 18th Ave., Columbus, OH 43210, USA                                                   \label{inst10}  
\and  School of Space Research, Kyung Hee University, Yongin, Kyeonggi 17104, Republic of Korea                                                       \label{inst31}  
\and  Department of Particle Physics and Astrophysics, Weizmann Institute of Science, Rehovot 76100, Israel                                           \label{inst11}  
\and  Department of Physics, University of Warwick, Gibbet Hill Road, Coventry CV4 7AL, UK                                                            \label{inst12}  
\and  Villanova University, Department of Astrophysics and Planetary Sciences, 800 Lancaster Ave., Villanova, PA 19085, USA                           \label{inst13}  
\and  Institute for Space-Earth Environmental Research, Nagoya University, Nagoya 464-8601, Japan                                                     \label{inst14}  
\and  Code 667, NASA Goddard Space Flight Center, Greenbelt, MD 20771, USA                                                                            \label{inst15}  
\and  Department of Astronomy, University of Maryland, College Park, MD 20742, USA                                                                    \label{inst16}  
\and  Department of Earth and Planetary Science, Graduate School of Science, The University of Tokyo, 7-3-1 Hongo, Bunkyo-ku, Tokyo 113-0033, Japan   \label{inst17}  
\and  Instituto de Astrof{\'i}sica de Canarias, V{\'i}a L{\'a}ctea s/n, E-38205 La Laguna, Tenerife, Spain                                            \label{inst18}  
\and  Department of Earth and Space Science, Graduate School of Science, Osaka University, Toyonaka, Osaka 560-0043, Japan                            \label{inst19}  
\and  Institute of Astronomy, Graduate School of Science, The University of Tokyo, 2-21-1 Osawa, Mitaka, Tokyo 181-0015, Japan                        \label{inst25}  
\and  Oak Ridge Associated Universities, Oak Ridge, TN 37830, USA                                                                                     \label{inst20}  
\and  Institute of Space and Astronautical Science, Japan Aerospace Exploration Agency, 3-1-1 Yoshinodai, Chuo, Sagamihara, Kanagawa 252-5210, Japan  \label{inst26}  
\and  Sorbonne Universit\'e, CNRS, UMR 7095, Institut d'Astrophysique de Paris, 98 bis bd Arago, 75014 Paris, France                                  \label{inst22}  
\and  Department of Physics, University of Auckland, Private Bag 92019, Auckland, New Zealand                                                         \label{inst23}  
\and  University of Canterbury Mt.~John Observatory, P.O. Box 56, Lake Tekapo 8770, New Zealand                                                       \label{inst24}  
\and  Department of Astronomy, Graduate School of Science, The University of Tokyo, 7-3-1 Hongo, Bunkyo-ku, Tokyo 113-0033, Japan                     \label{inst21}  
}                                                                                                                                                                     
\date{Received ; accepted}

\abstract
{}
{
The United Kingdom Infrared Telescope (UKIRT) microlensing survey was conducted over four 
years, from 2016 to 2019, with the goal of serving as a precursor to future near-infrared 
microlensing surveys \citep{Shvartzvald2017}.  Focusing on stars in the Galactic center and 
utilizing near-infrared passbands, the survey identified approximately one thousand microlensing 
events, 27 of which displayed anomalies in their light curves \citep{Wen2023}. This paper 
presents an analysis of these anomalous events, aiming to uncover the underlying causes of 
the observed anomalies.
}
{
The events were analyzed under various configurations, considering the potential binarity of 
both the lens and the source. For 11 events that were additionally observed by other optical 
microlensing surveys, including those conducted by the OGLE, KMTNet, and MOA collaborations, 
we incorporated their data into our analysis.
}
{
Among the reported anomalous events, we revealed the nature of 24 events except for three events, 
in which one was likely to be a transient variable, and two were were difficult to accurately 
characterize their nature due to the limitations of the available data.  We confirmed the binary 
lens nature of the anomalies in 22 events.  Among these, we verified the earlier discovery that 
the companion in the binary lens system UKIRT11L is a planetary object.  Accurately describing 
the anomaly in UKIRT21 required a model that accounted for the binarity of both the lens and the 
source.  For two events UKIRT01 and UKIRT17, the anomalies could be interpreted using either a 
binary-source or a binary-lens model.  For the UKIRT05, it was found that accounting for higher-order 
effects induced by the orbit al motions of both Earth and the binary lens was crucial.  With the 
measured microlensing parallax togeter with the angular Einstein radius, the component masses of 
the UKIRT05 binary lens were determined to be $M_1 = (1.05 \pm 0.20)~M_\odot$, $M_2 = (0.36 \pm 
0.07)~M_\odot$, and the distance to the lens was found to be $\dl = (3.11 \pm 0.40)$~kpc.
}
{}

\keywords{gravitational lensing: micro}

\maketitle

\begin{table*}[t]
\small
\caption{Event coordinates and correspondence.  \label{table:one}}
\begin{tabular}{ccccccc}
\hline\hline
\multicolumn{1}{c}{UKIRT}                      &
\multicolumn{1}{c}{(RA, DEC)$_{\rm J2000}$}    &
\multicolumn{1}{c}{OGLE}                       &
\multicolumn{1}{c}{KMTNet}                     &
\multicolumn{1}{c}{MOA}                        \\
\hline
  UKIRT-01   & (17:54:25.15, -27:42:01.6)    &  --                  &  --                  &   --                 \\
  UKIRT-02   & (18:05:34.12, -27:39:39.1)    &  OGLE-2016-0BLG-562  &  KMT-2016-BLG-0042   &   --                 \\
  UKIRT-03   & (17:58:17.28, -28:40:52.2)    &  --                  &  --                  &   --                 \\
  UKIRT-04   & (17:58:33.87, -27:22:43.6)    &  OGLE-2016-BLG-0887  &  KMT-2016-BLG-0617   &   --                 \\
  UKIRT-05   & (17:54:50.52, -28:19:07.1)    &  --                  &  KMT-2016-BLG-0180   &   --                 \\
  UKIRT-06   & (17:50:20.87, -29:18:55.6)    &  OGLE-2016-BLG-0262  &  KMT-2016-BLG-0592   &   --                 \\
  UKIRT-07   & (17:53:11.35, -27:42:32.6)    &  --                  &  --                  &   --                 \\
  UKIRT-08   & (17:42:45.92, -29:13:15.1)    &  --                  &  --                  &   --                 \\
  UKIRT-09   & (17:47:17.25, -29:38:43.4)    &  --                  &  --                  &   --                 \\
  UKIRT-10   & (17:49:10.47, -29:39:20.4)    &  --                  &  --                  &   --                 \\
  UKIRT-11   & (17:46:36.98, -29:12:40.9)    &  --                  &  --                  &   --                 \\
  UKIRT-12   & (17:49:08.63, -27:57:09.5)    &  --                  &  --                  &   --                 \\
  UKIRT-13   & (17:54:08.71, -27:45:16.9)    &  --                  &  --                  &   --                 \\
  UKIRT-14   & (17:54:58.96, -28:26:21.8)    &  OGLE-2017-BLG-1323  &  KMT-2017-BLG-0213   &   MOA-2017-BLG-431   \\
  UKIRT-15   & (17:51:24.11, -29:17:27.7)    &  OGLE-2017-BLG-1471  &  KMT-2017-BLG-0318   &   MOA-2017-BLG-423   \\
  UKIRT-16   & (17:45:20.79, -28:12:43.7)    &  --                  &  --                  &   --                 \\
  UKIRT-17   & (17:50:58.26, -27:49:00.1)    &  --                  &  --                  &   --                 \\
  UKIRT-18   & (17:52:28.43, -28:23:35.4)    &  OGLE-2018-BLG-0752  &  postseason          &   --                 \\
  UKIRT-19   & (17:52:37.79, -28:36:14.1)    &  OGLE-2018-BLG-1055  &  KMT-2018-BLG-2095   &   --                 \\
  UKIRT-20   & (17:51:46.70, -28:12:42.1)    &  OGLE-2018-BLG-0856  &  KMT-2018-BLG-2392   &   --                 \\
  UKIRT-21   & (17:41:49.39, -29:40:15.7)    &  --                  &  --                  &   --                 \\
  UKIRT-22   & (17:42:36.48, -29:30:21.9)    &  --                  &  --                  &   --                 \\
  UKIRT-23   & (17:42:39.02, -28:59:40.8)    &  --                  &  --                  &   --                 \\
  UKIRT-24   & (17:41:27.98, -29:00:02.8)    &  --                  &  --                  &   --                 \\
  UKIRT-25   & (17:50:08.93, -28:19:52.9)    &  --                  &  --                  &   --                 \\
  UKIRT-26   & (17:55:44.43, -29:45:10.5)    &  OGLE-2019-BLG-0950  &  KMT-2019-BLG-1326   &   MOA-2019-BLG-277   \\
  UKIRT-27   & (17:54:39.34, -29:06:14.7)    &  OGLE-2019-BLG-1048  &  KMT-2019-BLG-1450   &   MOA-2019-BLG-363   \\
\hline
\end{tabular}
\end{table*}

\section{Introduction} \label{sec:one}

A microlensing survey was conducted over four years, from 2016 to 2019, using the United
Kingdom Infrared Telescope (UKIRT) to observe stars in the Galactic center in the near-infrared
(NIR) passband \citep{Shvartzvald2017}.  The survey targeted the
Galactic center field, reaching $l = b = 0^\circ$, a region difficult to observe with optical 
lensing surveys such as the Korea Microlensing Telescope Network \citep[KMTNet;][]{Kim2016}, 
the Optical Gravitational Lensing Experiment \citep[OGLE;][]{Udalski2015}, and the Microlensing 
Observations in Astrophysics \citep[MOA;][]{Bond2001, Sumi2003}. The primary objective of the 
UKIRT survey was to examine the spatial variation in microlensing event rates across the Galactic 
center field. The results from this survey provide valuable insights to refine and optimize future 
NIR microlensing surveys, including those planned for the Nancy Grace Roman Space Telescope 
\citep{Spergel2015}, by enhancing predictions and detections of microlensing events in densely 
populated, dust-obscured regions of the sky.

During the four-year survey period, the UKIRT microlensing survey identified a total of 985
microlensing events. Of these, 522 were definitively classified as single-lens single-source 
(1L1S) events, where the flux of a single source star was gravitationally magnified by a 
single-mass lens.  An additional 436 events were categorized as probable 1L1S events, which 
are considered likely microlensing events despite not meeting all classification criteria.  
Detailed lensing parameters for these 1L1S events are provided in Table A1 of \citet{Wen2023}.

In addition to the standard 1L1S events, \citet{Wen2023} identified 27 candidate anomalous 
events with light curves that deviated from the smooth and symmetric profiles typical of 1L1S 
events.  These deviations suggest more complex lensing configurations, but detailed analyses 
of these anomalous events have not yet been conducted, leaving their exact nature uncertain.

Some of these events are likely binary-lens single-source (2L1S) events, with the anomalies
resulting from the source star crossing the caustic structures created by the binary lens 
system.  Such caustic crossings typically generate sharp and distinct features in the light 
curve. For events without clear caustic-crossing signatures, their origins remain ambiguous, 
potentially involving alternative lensing configurations, such as single-lens binary-source 
(1L2S) systems. Further analysis will be essential to classify these events and uncover the 
physical mechanisms underlying their anomalous features.

In this paper, we  present detailed analyses of the anomalous microlensing events detected by 
the UKIRT microlensing survey.  We first examine the binary nature of both the lens and the 
source by modeling the light curves using binary-lens and binary-source configurations. If 
these models fail to fully account for the anomalies, we investigate more complex scenarios 
by incorporating additional lens or source components into the 2L1S and 1L2S configurations.

The paper is organized as follows. Section~\ref{sec:two} provides a detailed description 
of the data used in our analyses. For UKIRT events with additional data from other lensing 
surveys, we incorporated all available datasets to ensure a comprehensive analysis. A list 
of these events, along with the corresponding additional data, is also included for reference. 
In Section~\ref{sec:three}, we discuss the primary causes of anomalies in lensing light curves 
and outline the parameters required for modeling such anomalies.  Additionally, we explain 
the procedure of the lensing modeling conducted to find a set of lensing parameters that best 
describe observed light curves.  Section~\ref{sec:four} focuses on the modeling process for 
each individual anomalous lensing event. We describe the specific characteristics of the 
anomalies observed in each event, provide a detailed explanation of the modeling procedure, 
and present the results derived from the modeling.  In Section~\ref{sec:five}, we estimate 
the angular Einstein radii for events with light curves showing finite-source effects and 
available $I$- and $V$-band photometry.  We summarize results and conclude in 
Section~\ref{sec:six}.

\section{Data} \label{sec:two}

The UKIRT microlensing event data analyzed in this study were obtained from lensing survey 
conducted using the UKIRT telescope. The telescope has a 2.8-meter aperture and is 
equipped with the Wide Field Camera, featuring a pixel scale of 0.4" and a field of view 
of approximately 0.8~deg$^2$ \citep{Shvartzvald2017}. Observations were 
carried out in two near-infrared passbands, $H$ and $K$, with the data processed using the 
CASU multi-aperture photometry pipeline \citep{Irwin2004, Hodgkin2009}.  The UKIRT survey 
fields vary depending on the observation year but are consistently located near the Galactic 
center, within a Galactic latitude of $b<3^\circ$, extending to $l=b=0^\circ$. This design
serves the survey's primary objective of examining the spatial variation in microlensing 
event rates across the Galactic center observation field.

\begin{figure}[t]
\includegraphics[width=\columnwidth]{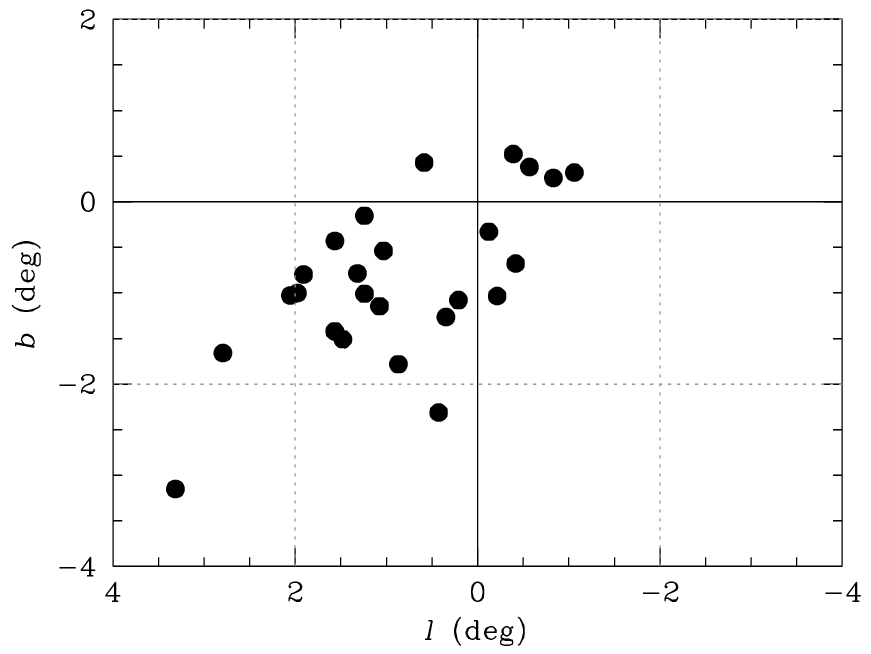}
\caption{
Positions of anomalous events found by the UKIRT survey in Galactic coordinates.
}
\label{fig:one}
\end{figure}

Among the 27 anomalous transients with light curves deviating from the 1L1S lensing form, 
26 were identified as lensing events, with the exception of UKIRT03, which is likely a 
variable star. The positions of these anomalous UKIRT events in Galactic coordinates are 
shown in Figure~\ref{fig:one}, while their Equatorial coordinates are provided in Paper~I. 
Of the reported anomalous events, 11 were also observed by other optical microlensing surveys 
conducted by the OGLE, KMTNet, and MOA groups. Table~\ref{table:one} lists the events observed 
by multiple surveys, along with the identification references assigned by each survey group. 
In this work, we use the UKIRT references to designate the events.

Data from the other lensing surveys were obtained using their respective instruments. The 
OGLE survey is carried out using a 1.6-meter telescope equipped with a camera covering a 1.4
square-degree field, located at Las Campanas Observatory in Chile. The KMTNet survey employs 
a network of three identical 1.6-meter telescopes situated at geographically distributed sites: 
Siding Spring Observatory in Australia (KMTA), Cerro Tololo Inter-American Observatory in Chile
(KMTC), and the South African Astronomical Observatory in South Africa (KMTS). Each KMTNet
telescope is fitted with a camera that provides a 4 square-degree field of view. Meanwhile, t
he MOA survey operates a 1.8-meter telescope at Mt. John University Observatory in New Zealand,
equipped with a camera covering a 2.2 square-degree field. Observations from OGLE and KMTNet
were primarily performed in the $I$-band, while the MOA survey utilized a custom MOA-$R$
band, spanning wavelengths from 609 to 1109 nm. The data from these additional surveys were
reduced using photometry pipelines specific to each survey group: the OGLE data were processed
with the pipeline developed by \citet{Udalski2003}, the KMTNet data were analyzed using the 
method described by \citet{Albrow2009}, and the MOA data were reduced following the approach
outlined by \citet{Bond2001}. In the modeling process, we adjusted the error bars of the data 
to match the data scatter and to ensure that the $\chi^2$ per degree of freedom for each data 
set equals unity, following the procedure outlined in \citet{Yee2012}.

\section{Lensing model} \label{sec:three}

The lensing magnification of a 1L1S event depends on the projected separation between the lens 
and the source, normalized to the angular Einstein radius ($\thetae$), following the relation:
\begin{equation}
A = {u^2+2 \over u(u^2+4)^{1/2}};\qquad
u = \left[ u_0^2 + \left( {t-t_0 \over \te}\right)^2\right]^{1/2}.
\label{eq1}
\end{equation}
Here, $t_0$ represents the time at the closest approach between the lens and the source, and 
$u_0$ (normalized to $\thetae$) indicates the separation at $t_0$ (impact parameter).  The 
parameter $\te$ denotes the event time scale, defined as the time it takes the source to traverse 
the Einstein ring radius.  As a result, modeling the light curve of a 1L1S event requires three 
lensing parameters: $(t_0, u_0, \te)$. The resulting lensing light curve is smooth and symmetric 
about $t_0$ \citep{Paczynski1986}.

A lensing light curve may deviate from the standard 1L1S form due to various factors, with the 
binarity of the lens being the most common cause \citep{Mao1991}. In 2L1S events, the lensing 
system creates a caustic structure on the source plane, which represents locations at which the 
lensing magnification of a point source becomes infinite. As a result, 2L1S light curves show 
deviations from the 1L1S form when the source approaches or crosses the caustic.  The shape, 
size, and position of the caustic vary depending on the projected separation between the lens 
components ($s$, normalized to $\thetae$) and their mass ratio ($q$) \citep{Cassan2008}.  
Anomalies caused by caustic crossings are affected by finite-source effects. In the case of 
an extended source event, the observed magnification is determined by the intensity-weighted 
amplification averaged over the surface of the source star \citep{Nemiroff1994, Witt1994, 
Bennett1996, Gould1997}. Consequently, finite-source effects smooth out the sharp caustic spikes 
in the lensing light curves.  The presence of a caustic structure introduces four additional 
parameters, $(s, q, \alpha, \rho)$, necessary for modeling a 2L1S event. Here, $\alpha$ denotes 
the angle between the source motion direction and the binary-lens axis, while $\rho$ represents 
the ratio of the angular source radius ($\theta_*$) to $\thetae$ (normalized source radius).

Another important cause of deviations from the 1L1S form is the binarity of the source.  In a 
1L2S event, the lensing light curve is the combined result of the contributions from the events 
associated with each individual source star \citep{Griest1992, Stefano1995, Han1997, Dominik1998}.  
As a result, the light curve exhibits deviations from the smooth, single-peak profile characteristic 
of a 1L1S event. Modeling the light curve of 1L2S events requires three additional parameters to 
describe the source companion: $(t_{0,2}, u_{0,2}, q_F)$. Here, $(t_{0,2}, u_{0,2})$ represent 
the impact parameter and approach time of the second source, while $q_F$ denotes the flux ratio 
between the secondary and primary source stars.  If the light curve is influenced by the finite 
size of the second source, an additional parameter, $\rho_2$, representing the normalized radius 
of the second source, must be included in the modeling.

\begin{table*}[t]
\caption{Lensing parameters of UKIRT01.  \label{table:two}}
\begin{tabular}{lllllll}
\hline\hline
\multicolumn{1}{c}{Parameter}   &
\multicolumn{2}{c}{2L1S}      &
\multicolumn{1}{c}{1L2S}      \\
\multicolumn{1}{c}{}   &
\multicolumn{1}{c}{Close}      &
\multicolumn{1}{c}{Wide}      &
\multicolumn{1}{c}{}      \\
\hline
 $\chi^2$                   &   89.5                &   89.8                 &   89.5                    \\
 $t_0$ (HJD$^\prime$)       &  $7522.07 \pm 0.12$   &  $7521.88 \pm 0.12$    &  $7521.225 \pm 0.255$     \\
 $u_0$                      &  $0.123 \pm 0.043 $   &  $0.111 \pm 0.027 $    &  $0.193 \pm 0.097   $     \\
 $\te$ (days)               &  $21.05 \pm 6.39  $   &  $27.79 \pm 9.67  $    &  $24.17 \pm 6.86    $     \\
 $s$                        &  $0.438 \pm 0.082 $   &  $2.604 \pm 0.813 $    &  --                       \\
 $q$                        &  $0.33 \pm 0.16   $   &  $0.19 \pm 0.69   $    &  --                       \\
 $\alpha$ (rad)             &  $4.489 \pm 0.055 $   &  $4.459 \pm 0.061 $    &  --                       \\
 $\rho$ ($10^{-3}$)         &  --                   &  --                    &  --                       \\
 $t_{0,2}$ (HJD$^\prime$)   &  --                   &  --                    &  $7522.700 \pm 0.119$     \\
 $u_{0,2}$ (HJD$^\prime$)   &  --                   &  --                    &  $0.015 \pm 0.023   $     \\
 $q_F$                      &  --                   &  --                    &  $0.159 \pm 0.050   $     \\
\hline
\end{tabular}
\end{table*}

We initially model the anomalous UKIRT events using the most common lens configurations, 2L1S
and 1L2S systems.  The goal of the modeling is to identify the set of lensing parameters (lensing
solution) that best describe the observed light curves. For the 2L1S system, we first search for the
binary parameters $(s, q)$ using a grid approach with multiple starting values of $\alpha$, while
the other parameters are explored through a downhill method. Subsequently, we refine the solution
by allowing all parameters to vary. For the 1L2S configuration, we determine the lensing solution
by considering the location and magnitude of the anomaly in the light curve.

In some rare instances where anomalies in lensing light curves are challenging to describe
precisely with a 2L1S or a 1L2S model, we employed more complex modeling.  We consider two
scenarios, with the first being a 3L1S configuration where the lens is a triple system and the
source is a single star, as exemplified by the lensing event KMT-2021-BLG-1122 \citep{Han2023}, 
and the second being a 2L2S configuration where both the lens and the source are binaries, as
demonstrated in the events KMT-2021-BLG-0284, KMT-2022-BLG-2480, and KMT-2024-BLG-0412
\citep{Han2024a}.

For events with long time scales and well-sampled light curves, we conducted additional modeling 
to account for higher-order effects that could impact the light curve.  Specifically, we considered 
two effects: microlens parallax and lens orbital motion. The microlens-parallax effect arises from 
the movement of Earth in its orbit around the Sun, which leads to a shift in the observer’s position 
\citep{Gould1992}. On the other hand, the lens-orbital effect is caused by the motion of the binary 
lens components in orbit around their common center of mass, which also results in a positional 
change of the lens system \citep{Batista2011, Skowron2011}. The modeling considering the 
microlens-parallax effect requires two additional parameters, $(\pien, \piee)$, which correspond 
to the north and east components of the microlens-parallax vector $\pivec_{\rm E}$. The direction 
of the microlens-parallax vector aligns with that of the relative lens-source proper motion vector 
($\muvec$), and its magnitude is related to the relative lens-source parallax ($\pi_{\rm rel}$) by 
$\pie = \pi_{\rm rel}/\thetae$. To first-order approximation, the lens orbital motion is described 
by two parameters, $(ds/dt, d\alpha/dt)$, which represent the annual rates of change in the binary 
separation and the source trajectory angle, respectively \citep{Albrow2000, Jung2013}.

\section{Lens-system configuration} \label{sec:four}

In this section, we present the results of modeling the individual anomalous UKIRT lensing 
events.  For each event, we provide the lensing solution and describe the corresponding 
lens-system configuration. In cases where multiple solutions arise due to degeneracies, we 
present all possible solutions and explain the cause of the degeneracy. Lensing solutions 
are provided for 24 of the 26 anomalous UKIRT lensing events, with the exception of UKIRT10 
and UKIRT24, for which accurate characterization is difficult due to the limitations of the 
available data.

\begin{figure}[t]
\includegraphics[width=\columnwidth]{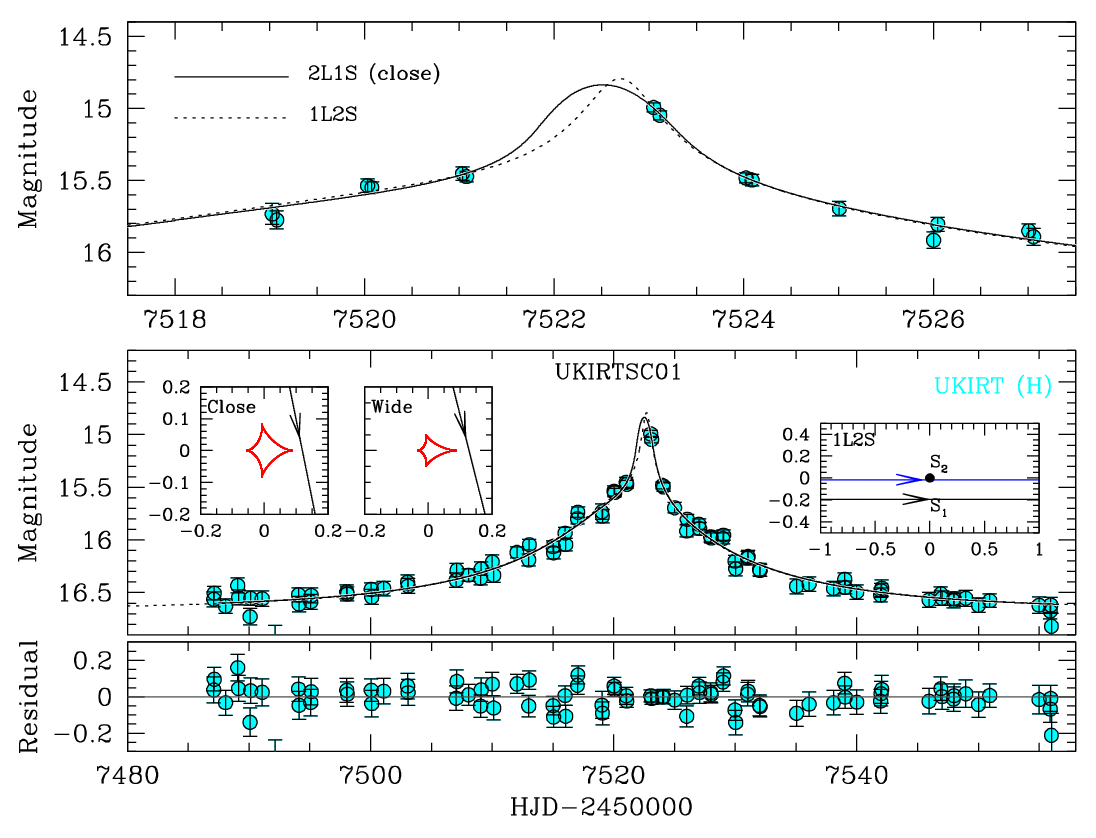}
\caption{
The lensing light curve of UKIRT01.  The lower panel displays the full view of the
light curve, while the upper panel provides a zoomed-in view of the anomalous region. The
insets in the lower panel illustrate the lens system configurations for the 2L1S model (left
two insets) and the 1L2S solution (right inset).  In the 2L1S configuration, the red closed curve
represents the caustic, and the arrowed line indicates the source trajectory. In the 1L2S
solution, the lens position is marked by a filled dot at the origin, while the blue and black
arrowed lines represent the trajectories of the primary ($S_1$) and secondary ($S_2$) source 
stars, respectively.
}
\label{fig:two}
\end{figure}

\subsection{UKIRT01} \label{sec:four-one}

Figure~\ref{fig:two} presents the lensing light curve of UKIRT01, which occurred in the  
2016 observing season. This source of the event, located at Galactic coordinates $(l, b) = 
(2^\circ\hskip-2pt .0545, -1^\circ\hskip-2pt .0277)$, was exclusively monitored by the UKIRT 
survey, as it lay outside the coverage areas of other surveys. Observations during the 2016 
season were conducted in a single passband, $H$, with two data points collected daily to 
track the event. The light curve peaked around May 12, 2016, corresponding to the reduced 
heliocentric Julian date $(\hjd^\prime \equiv \hjd - 2450000 \sim 7522.2$.

The two data points recorded on $\hjd^\prime = 7523$  exhibit deviations from the 1L1S form. 
Modeling the event with a 2L1S configuration revealed two solutions arising from the well-known 
close-wide degeneracy. In this degeneracy, one solution corresponds to a binary separation 
greater than unity $(s > 1$, "wide" solution), while the other corresponds to a separation 
less than unity ($s < 1$, "close" solution).  The degeneracy between these solutions is 
notable, with $\Delta\chi^2 = 0.3$.  Figure~\ref{fig:two} displays the model curve for the 
close solution along with its residuals.  The wide solution exhibits a very similar profile. 
The lens system configurations, showing the source trajectory relative to the caustic produced 
by the binary lens, are illustrated in the inner two insets of the lower panel. The full lensing 
parameters for both 2L1S solutions are listed in Table~\ref{table:two}.  The estimated mass 
ratio between the binary lens components is $q \sim 0.44$ for the close solution and $q \sim 
0.19$ for the wide solution. In both cases, the anomaly is attributed to the source's passage 
through the extended perturbation region created by the protruding central caustic.

It is found that the anomaly can also be explained by a model with a 1L2S configuration. The
model curve for the 1L2S solution is shown as a dotted line in Figure~\ref{fig:two}, and the 
corresponding model parameters are listed in Table~\ref{table:two}. Under the 1L2S interpretation, 
the anomaly arises from the close approach of the faint companion of the source to the lens. 
The estimated flux ratio between the source companion and the primary source is $q_F \sim 0.16$. 
The lens system configuration is shown in the right inset of the lower panel.

\begin{table*}[t]
\caption{Best-fit parameters of UKIRT02, UKIRT04, UKIRT05, UKIRT06, and UKIRT07.  \label{table:three}}
\begin{tabular}{lcccccc}
\hline\hline
\multicolumn{1}{c}{Parameter}    &
\multicolumn{1}{c}{UKIRT02}      &
\multicolumn{1}{c}{UKIRT04}      &
\multicolumn{1}{c}{UKIRT05}      &
\multicolumn{1}{c}{UKIRT06}      &
\multicolumn{1}{c}{UKIRT07}      \\
\hline
 $\chi^2$             &  6685.6               &  7472.2               &  6497.8               &  7722.9                &  136.9               \\
 $t_0$ (HJD$^\prime$) &  $7492.893 \pm 0.012$ &  $7536.53 \pm 0.21 $  &  $7548.00 \pm 0.22 $  &  $7515.929 \pm 0.013$  &  $7556.38 \pm 0.45 $ \\
 $u_0$                &  $0.1508 \pm 0.0025 $ &  $0.8232 \pm 0.0073$  &  $0.0628 \pm 0.0017$  &  $0.1229 \pm 0.0019 $  &  $0.0952 \pm 0.0088$ \\
 $\te$ (days)         &  $20.64 \pm 0.12    $ &  $27.20 \pm 0.13   $  &  $93.98 \pm 0.21   $  &  $74.66 \pm 1.37    $  &  $104.96 \pm 6.10  $ \\
 $s$                  &  $1.6593 \pm 0.0018 $ &  $1.6251 \pm 0.0048$  &  $1.5272 \pm 0.0021$  &  $4.421 \pm 0.056   $  &  $0.981 \pm 0.023  $ \\
 $q$                  &  $0.982 \pm 0.022   $ &  $0.374 \pm 0.012  $  &  $0.3320 \pm 0.0025$  &  $1.114 \pm 0.093   $  &  $0.238 \pm 0.024  $ \\
 $\alpha$ (rad)       &  $4.6733 \pm 0.0050 $ &  $5.0613 \pm 0.0081$  &  $3.1955 \pm 0.0028$  &  $5.7136 \pm 0.0019 $  &  $0.058 \pm 0.022  $ \\
 $\rho$ ($10^{-3}$)   &  $0.957 \pm 0.014   $ &  $0.996 \pm 0.016  $  &  $0.5780 \pm 0.0043$  &   --                   &  $2.98 \pm 0.49    $ \\
 $\pien$              &   --                  &   --                  &  $-0.091 \pm 0.010 $  &   --                   &    --                \\
 $\piee$              &   --                  &   --                  &  $0.068 \pm 0.008  $  &   --                   &    --                \\
 $ds/dt$              &   --                  &   --                  &  $0.099 \pm 0.025  $  &   --                   &    --                \\
 $d\alpha/dt$         &   --                  &   --                  &  $-0.429 \pm 0.022 $  &   --                   &    --                \\
\hline
\end{tabular}
\end{table*}

\subsection{UKIRT02} \label{sec:four-two}

The event UKIRT02 was detected by the UKIRT survey during the 2016 observation season and 
was also observed by the OGLE and KMTNet surveys, which designated it as OGLE-2016-BLG-0562 
and KMT-2016-BLG-0042, respectively. The event was located within KMTNet’s prime fields BLG03 
and BLG43, both observed with a 0.5-hour cadence, providing an effective combined cadence of 
0.25 hours. This resulted in a densely sampled light curve.

\begin{figure}[t]
\includegraphics[width=\columnwidth]{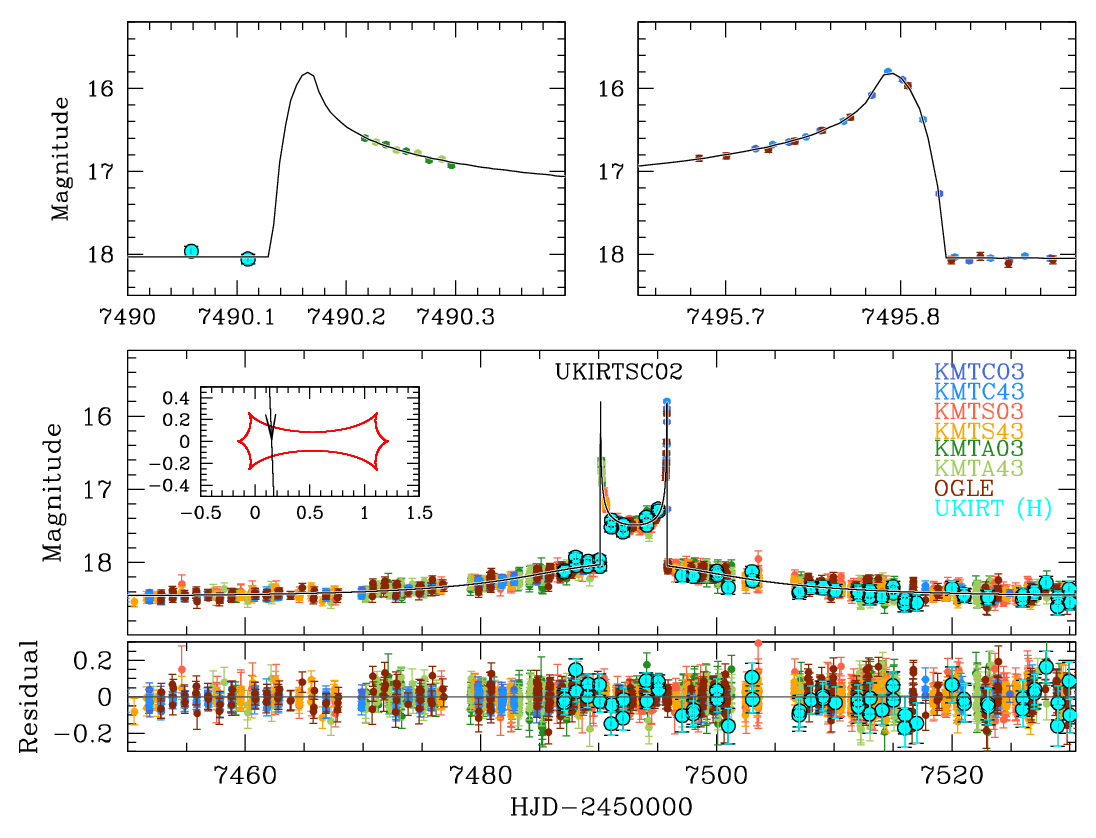}
\caption{
The lensing light curve of UKIRT02. The notations are the same as in Fig.~\ref{fig:two}.
}
\label{fig:three}
\end{figure}

Figure~\ref{fig:three} presents the lensing light curve of UKIRT02, which exhibits clear 
caustic-crossing features at $\hjd^\prime \sim 7490.1$ and 7495.8.  The second caustic 
crossing was resolved by a combination of the OGLE and KMTC data sets. From the 2L1S 
modeling, we identified a unique solution with binary-lens parameters $(s,q) \sim (1.65, 
0.98)$, indicating that the event was caused by a binary lens with nearly equal-mass 
components. The complete lensing parameters are provided in Table~\ref{table:three}, and 
the model curve is overlaid on the data points in Figure~\ref{fig:three}.

The inset in the lower panel illustrates the lens system configuration, revealing that 
the binary lens generated a single resonant caustic elongated along the binary axis. The 
source traversed the left side of the caustic nearly vertically, producing the first spike 
at the caustic entrance and the second spike at the exit. The measured event time scale is 
$\te \sim 21$ days, and the normalized source radius, $\rho \sim 0.96\times 10^{-3}$, was 
derived from the resolved caustic.

\subsection{UKIRT04} \label{sec:four-three}

The microlensing event UKIRT04 was the third anomalous event discovered by the UKIRT survey 
during the 2016 observing season. The source star of the event was located within the 
overlapping coverage regions of the OGLE and KMTNet surveys. The event IDs for these surveys 
are OGLE-2016-BLG-562 and KMT-2016-BLG-0042, respectively. Specifically, the source was 
situated in KMTNet's primary fields, BLG03 and BLG43, which were observed at a high cadence 
of 0.25 hours.

\begin{figure}[t]
\includegraphics[width=\columnwidth]{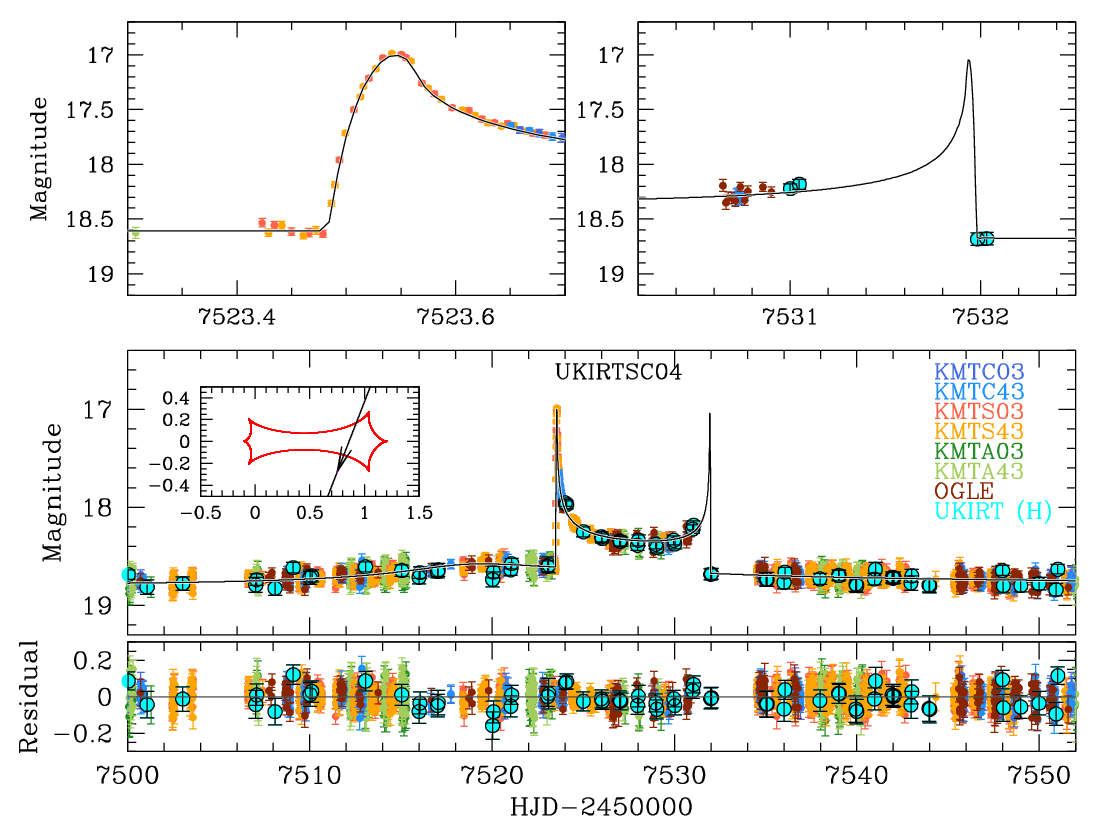}
\caption{
The lensing light curve of UKIRT04.  
}
\label{fig:four}
\end{figure}

Figure~\ref{fig:four} displays the light curve of UKIRT04, constructed by combining data 
from the three survey groups. Similar to UKIRT02, the light curve exhibits two distinct 
caustic spikes, with a U-shaped profile between them. The first spike, occurring at 
$\hjd^\prime\sim 7523.55$, was observed in the KMTC data set. Although the second caustic 
crossing was not directly resolved, it is estimated to have occurred at $\hjd^\prime\sim 
7531.95$, based on the rising slope of the U-shaped profile preceding the crossing. Modeling 
the event using the 2L1S configuration yields binary lens parameters $(s, q) \sim (1.63, 
0.37)$ and an event time scale of $\te \sim 27$ days. The normalized source radius, 
determined from the resolved caustic, was found to be $\rho \sim 1.00 \times 10^{-3}$. The 
full lensing parameters of the solution are listed in Table~\ref{table:three}.

The configuration of the lens system is illustrated in the inset of the lower panel. The 
caustic shape is similar to that of UKIRT02, with a single resonant caustic elongated along 
the binary lens axis. One key difference is that the source crosses the right side of the 
caustic at a diagonal angle.

\begin{figure}[t]
\includegraphics[width=\columnwidth]{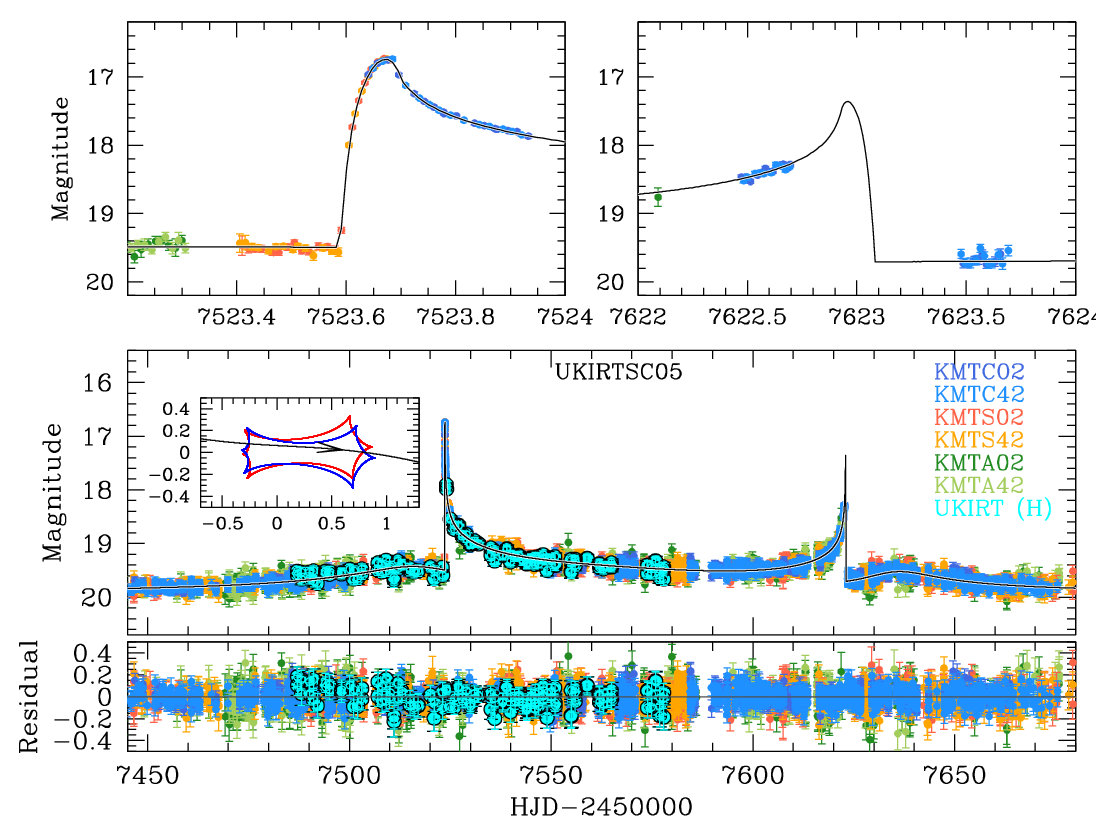}
\caption{
The lensing light curve of UKIRT05.  The two sets of caustics, depicted in red and blue 
in the inset of the lower panel, correspond to the caustics at the times of the first 
and second caustic crossings, respectively.
}
\label{fig:five}
\end{figure}

\subsection{UKIRT05} \label{sec:four-four}

The microlensing event UKIRT05, which occurred during the 2016 season, was observed by 
both the UKIRT and KMTNet surveys.  In the KMTNet survey, the event was designated as 
KMT-2016-BLG-0180 and monitored with a high cadence of 0.25 hours. This event persisted 
throughout the entire 2016 bulge season, with the UKIRT survey covering approximately 
the first half of the event. In contrast, the KMTNet survey provided continuous observations 
spanning the full duration of the bulge season, ensuring comprehensive coverage of the 
light curve.

Figure~\ref{fig:five} presents the lensing light curve of UKIRT05.  Similar to the two 
previous events, the light curve displays characteristic caustic crossing features, 
consisting of two spikes with a U-shaped trough between them. The first caustic spike, 
occurring at $\hjd^\prime \sim 7523.65$, was resolved by combined data from the KMTC and 
KMTS. Although the second spike was not directly observed, it is estimated to have occurred 
around $\hjd^\prime \sim 7622.9$, based on the light curve profile before and after the 
spike. In addition to these caustic spikes, the light curve shows a weak bump around 
$\hjd^\prime \sim 7645$, which appears after the second spike.

Taking into account the caustic features, we performed modeling using the 2L1S configuration.
Given the long duration of the event, we also included higher-order effects in the modeling. 
This approach yields a unique solution with binary parameters $(s,q)\sim (1.53,0.33)$. As 
expected from the event's extended duration, the estimated event time scale, $\te \sim 
94$~days, is notably long. The full set of lensing parameters for the solution is presented 
in Table~\ref{table:three}. The model fit that incorporates higher-order effects is found to 
be significantly better (by $\Delta\chi^2=354.5$) compared to the fit from the model without 
these effects.

The inset of the lower panel of Figure~\ref{fig:five} illustrates the configuration of the 
lens system.  Due to the orbital motion of the lens, the caustics change over time, and we 
present two sets of caustics corresponding to the times of the first (red closed curve) and 
second (blue closed curve) caustic spikes. The lens system produces a single resonant caustic 
that is elongated along the binary lens axis, with the source crossing through the longer 
side of the caustic. This, combined with the long time scale, results in a 100-day time gap 
between the two caustic spikes. The weak bump after the second caustic spike arouse as the 
source approached the right side online cusp of the caustic.

\begin{figure}[t]
\includegraphics[width=\columnwidth]{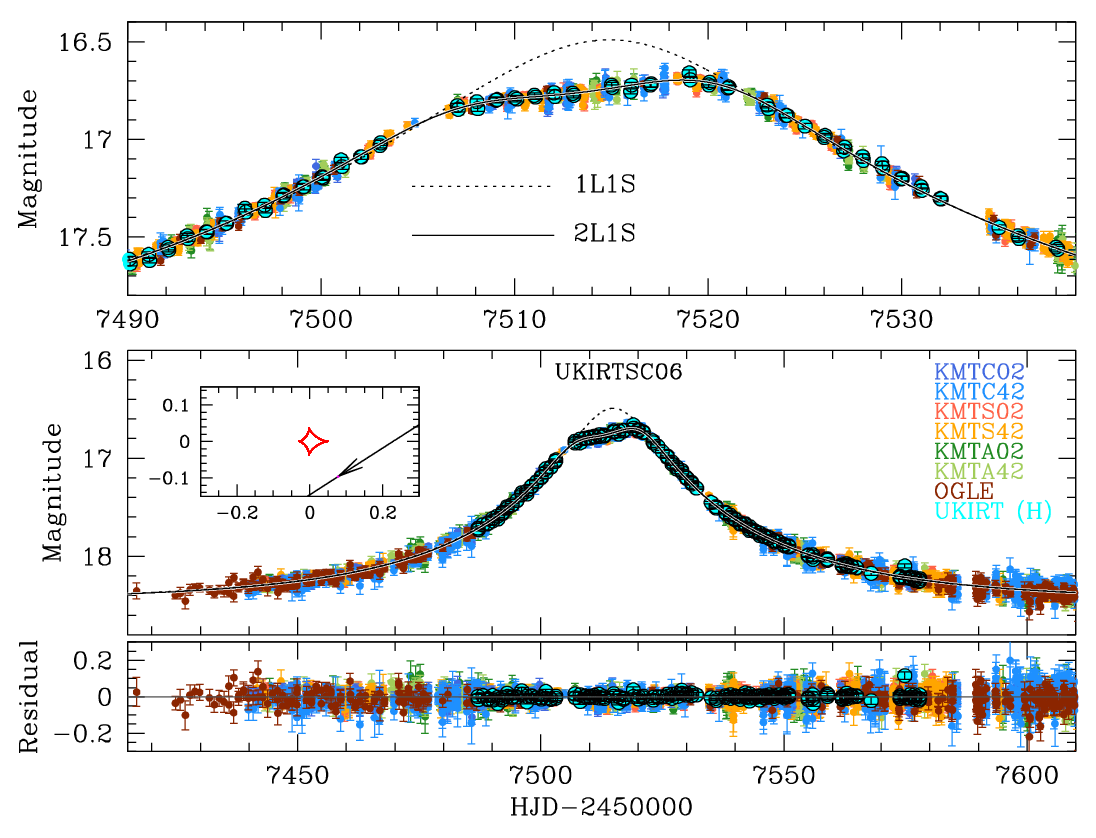}
\caption{
The lensing light curve of UKIRT06.  The dotted curve represents a 1L1S model
derived by fitting the data while excluding the data points around the anomaly.
}
\label{fig:six}
\end{figure}

\subsection{UKIRT06} \label{sec:four-five}

The event UKIRT06, occurred during the 2016 season, was observed not only by the UKIRT 
survey but also by two additional surveys conducted by the OGLE and KMTNet groups. 
Figure~\ref{fig:six} shows the light curve of the event, with the peak brightness 
occurring around $\hjd^\prime \sim 7516$, at which point the source star brightened by 
approximately 1.7 magnitudes above its baseline level. The light curve around the peak 
exhibited a subtle anomaly that persisted for about 17 days. This anomaly is characterized 
by a negative deviation from the 1L1S model and displays a nearly plateau-like shape during 
its duration.

\citet{Choi2012} suggested that a plateau-shaped central anomaly can be produced either 
by a planetary companion located near the Einstein ring of the primary lens or by a binary 
companion positioned well beyond the Einstein ring. Based on this, we performed 2L1S modeling, 
systematically exploring the parameter space for both the planetary and binary regimes. From 
this modeling, we derived a solution indicative of a binary lens system, with parameters 
$(s,q) \sim (4.4, 1.1)$. The complete set of lensing parameters for this solution is provided 
in Table~\ref{table:three}.

The model light curve corresponding to the solution is overlaid on the observed data points 
in Figure~\ref{fig:six}. The lens system configuration for this solution is illustrated in 
the inset of the lower panel. The configuration shows that the anomaly was caused by the 
source passing through the negative anomaly region located outside one fold of the caustic.

\begin{figure}[t]
\includegraphics[width=\columnwidth]{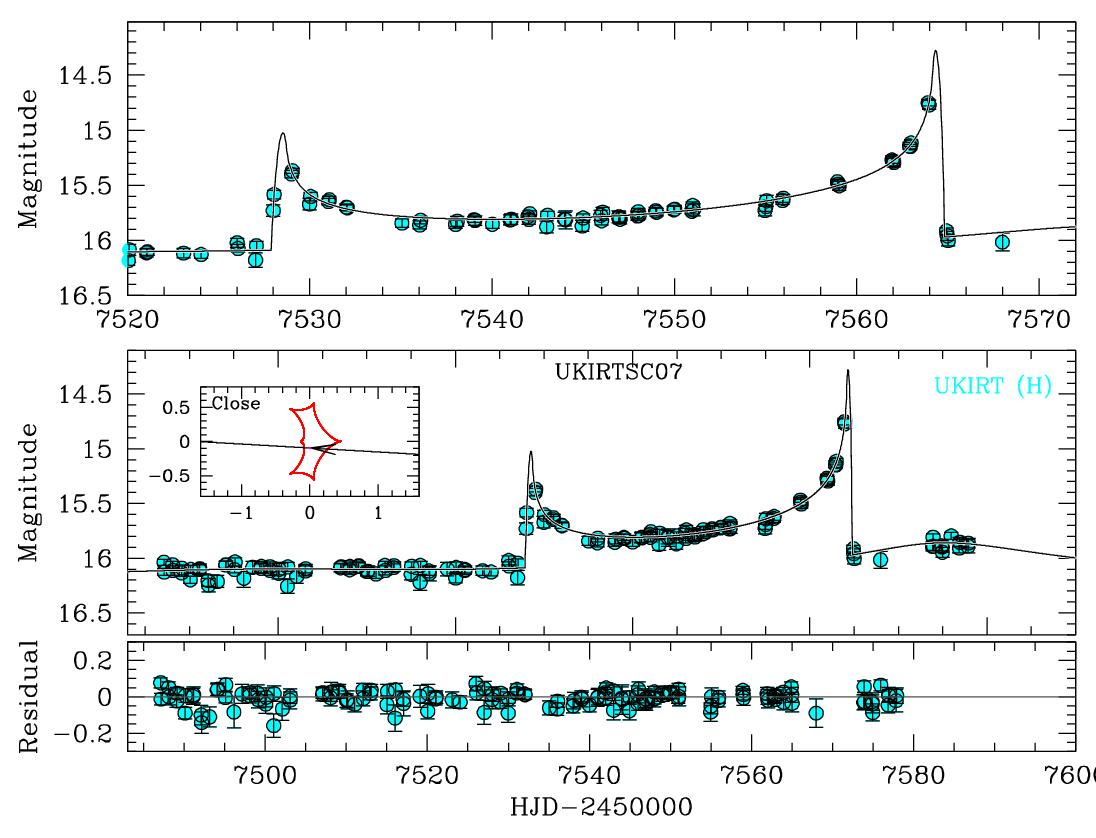}
\caption{
The lensing light curve of UKIRT07.  
}
\label{fig:seven}
\end{figure}

\subsection{UKIRT07} \label{sec:four-six}

The source of the event, located near the Galactic center at coordinates, $(l, b) = 
(1^\circ\hskip-2pt .9091, -0^\circ\hskip-2pt .7973)$, resides in a region not covered by other 
optical lensing surveys. Consequently, it was observed exclusively by the UKIRT survey
during the 2016 season.

The light curve of UKIRT07, shown in Figure~\ref{fig:seven}, clearly displays caustic-crossing 
features, with two prominent spikes: one occurring at $\hjd^\prime \sim 7528$ and the other at 
$\hjd^\prime \sim 7564$. Additionally, a weaker bump is observed after the second caustic spike, 
centered around $\hjd^\prime \sim 7585$. Despite the low cadence of the survey, the first caustic 
spike was partially covered, which allowed for the estimation of the normalized source radius.

\begin{table*}[t]
\caption{Best-fit parameters of UKIRT08, UKIRT09, UKIRT11, UKIRT12, and UKIRT13.  \label{table:four}}
\begin{tabular}{lcccccc}
\hline\hline
\multicolumn{1}{c}{Parameter}    &
\multicolumn{1}{c}{UKIRT08}      &
\multicolumn{1}{c}{UKIRT09}      &
\multicolumn{1}{c}{UKIRT11}      &
\multicolumn{1}{c}{UKIRT12}      &
\multicolumn{1}{c}{UKIRT13}      \\
\hline
 $\chi^2$             &  273.9              &  346.0                &  512.0                             &  36.1                  &  206.8                \\    
 $t_0$ (HJD$^\prime$) &  $7977.87 \pm 0.10$ &  $7954.31 \pm 7.14$   &  $7916.281 \pm 0.058            $  &  $7916.471 \pm 0.091$  &  $7981.653 \pm 0.080$ \\
 $u_0$                &  $0.325 \pm 0.032 $ &  $0.890 \pm 0.041 $   &  $4.01 \pm 0.22                 $  &  $4.68 \pm 0.56     $  &  $6.23 \pm 0.14     $ \\
 $\te$ (days)         &  $16.14 \pm 1.30  $ &  $103.64 \pm 9.63 $   &  $79.70 \pm 2.37                $  &  $40.25 \pm 1.59    $  &  $120.34 \pm 3.04   $ \\
 $s$                  &  $1.237 \pm 0.060 $ &  $1.559 \pm 0.046 $   &  $1.0144 \pm 0.0069             $  &  $3.72 \pm 0.13     $  &  $3.468 \pm 0.079   $ \\
 $q$                  &  $0.345 \pm 0.021 $ &  $0.101 \pm 0.087 $   &  $(1.88 \pm 0.18) \times 10^{-3}$  &  $2.38 \pm 0.84     $  &  $5.903 \pm 1.015   $ \\
 $\alpha$ (rad)       &  $1.462 \pm 0.019 $ &  $4.311 \pm 0.055 $   &  $0.579 \pm 0.018               $  &  $3.467 \pm 0.048   $  &  $-0.1951 \pm 0.0069$ \\
 $\rho$ ($10^{-3}$)   &  --                 &  --                   &  $9.10 \pm 0.80                 $  &  --                    &  --                   \\
\hline
\end{tabular}
\end{table*}

Modeling the light curve of the event using a 2L1S configuration yielded a solution with 
binary parameters of $(s, q)\sim (0.98, 0.24)$. The model curve of the solution is plotted 
in Figure~\ref{fig:seven}. The binary separation is very close to the angular Einstein 
radius, leading to the formation of a resonant caustic, as shown in the inset of the lower 
panel. The source crossed the caustic nearly horizontally along the binary lens axis. The 
weak bump following the second caustic spike was caused by the source’s close approach to 
the left-side on-axis cusp of the caustic. The event has a relatively long time scale of 
$\te \sim 105$ days, but determining higher-order lensing parameters was challenging due 
to the sparse and incomplete coverage of the light curve.

\subsection{UKIRT08} \label{sec:four-seven}

UKIRT08 was the first anomalous lensing event detected during the 2017 season by the UKIRT 
survey. Observations during this season were conducted in two passbands, $H$ and $K$. The 
source of UKIRT08 is located near the Galactic center, with Galactic coordinates $(l, b) 
\sim (-0^\circ\hskip-2pt .5691, 0^\circ\hskip-2pt .3823)$. This field lies outside the 
coverage of other optical lensing surveys, making the UKIRT survey the only one to observe 
the event. Observations for the season concluded at $\hjd^\prime = 7981$, before the event 
had fully returned to its baseline.

\begin{figure}[t]
\includegraphics[width=\columnwidth]{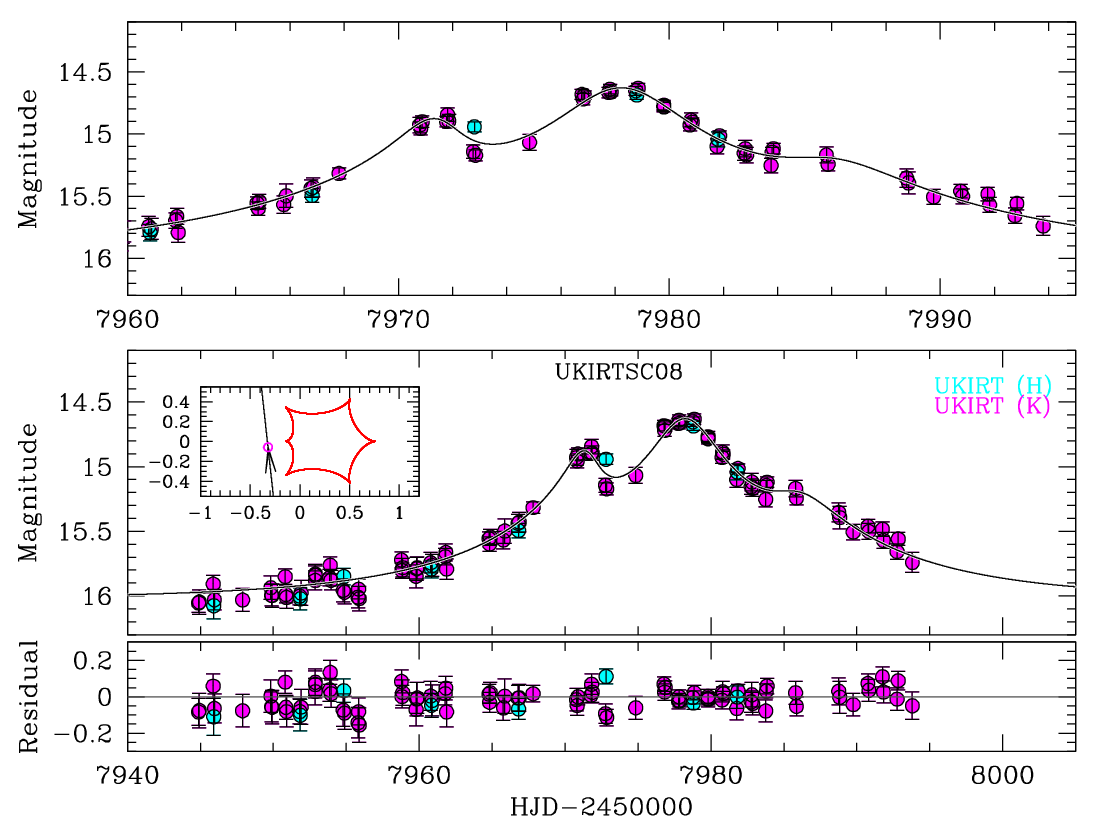}
\caption{
The lensing light curve of UKIRT08.  
}
\label{fig:eight}
\end{figure}

Figure~\ref{fig:eight} shows the lensing light curve of UKIRT08.  While it does not display 
caustic-crossing features, it is characterized by three consecutive bumps, centered at 
$\hjd^\prime \sim 7971$, 7979, and 7986. This anomaly pattern closely resembles that of 
the 2L1S lensing event KMT-2023-BLG-0601, identified through the KMTNet lensing survey 
\citep{Han2024b}. In the KMT-2023-BLG-0601 event, three closely spaced bumps were observed 
as the source passed nearly vertically along one side of a caustic, approaching three 
successive cusps.  This suggests that an anomaly characterized by three consecutive bumps 
is likely caused by a binary lens.

Given the likely binary-lens origin of the anomaly, we conducted a 2L1S modeling of the light 
curve. This analysis resulted in a solution with binary parameters $(s, q) \sim (1.24, 0.35)$. 
The lens system configuration, displayed in the inset of the lower panel, shows the formation 
of a single resonant caustic. As anticipated, based on the previous case with a similar anomaly 
pattern, the three anomaly features were caused by the vertical passage of the source through 
one side of the caustic. Table~\ref{table:four} provides the complete set of lensing parameters 
for the solution, excluding the normalized source radius, which could not be constrained.

\begin{figure}[t]
\includegraphics[width=\columnwidth]{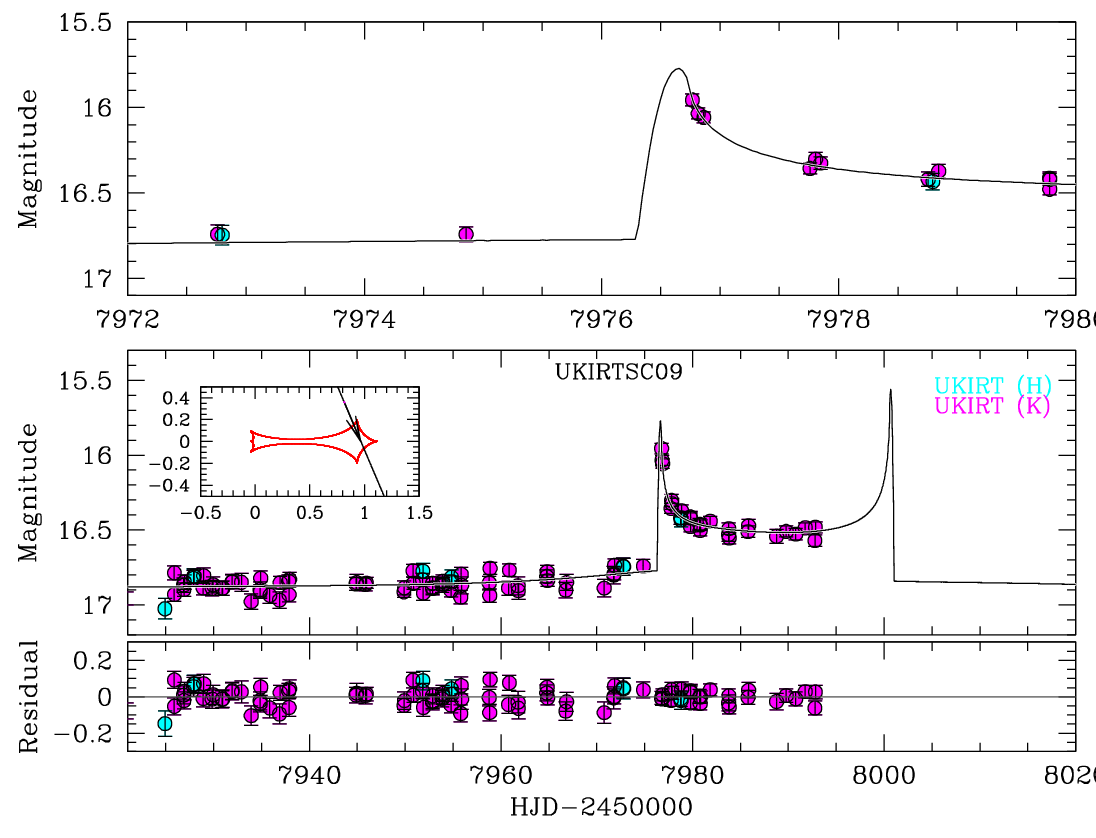}
\caption{
The lensing light curve of UKIRT09. 
}
\label{fig:nine}
\end{figure}

\subsection{UKIRT09} \label{sec:four-eight}

The lensing event UKIRT09, with its source located at Galactic coordinates $(l, b) = 
(-0^\circ\hskip-2pt .4171, -0^\circ\hskip-2pt .6791)$, was observed exclusively by the UKIRT 
survey. Figure~\ref{fig:nine} shows the light curve of UKIRT09, observed during the 2017 season. 
It features a prominent caustic spike at $\hjd^\prime = 7976.5$, marking the source’s entry into 
the caustic. However, the expected spike at the caustic exit was not observed, as it occurred 
after the observing season concluded, at $\hjd^\prime = 7989$.

Despite the incomplete coverage of the light curve, a 2L1S modeling provided a unique solution 
with binary parameters of approximately $(s, q) \sim (1.56, 0.10)$ and a time scale of $\te \sim 
104$ days.  The complet set of the lensing parameters are provided in Table~\ref{table:four}.

The lower panel of Figure~\ref{fig:nine} includes an inset showing the lens system configuration. 
The caustic structure consists of two distinct components: a smaller caustic near the more massive 
lens component and a larger caustic on the side of the lower-mass lens component, separated  by 
approximately $\Delta u \sim 0.9$. These two caustics are connected by a narrow bridge. The caustic 
spike occurred as the source passed through the larger caustic associated with the lighter lens 
component. According to the model, another spike was expected at $\hjd^\prime \sim 8001$, roughly 
12 days after the observing season concluded.

\subsection{UKIRT11} \label{sec:four-nine}

\begin{figure}[t]
\includegraphics[width=\columnwidth]{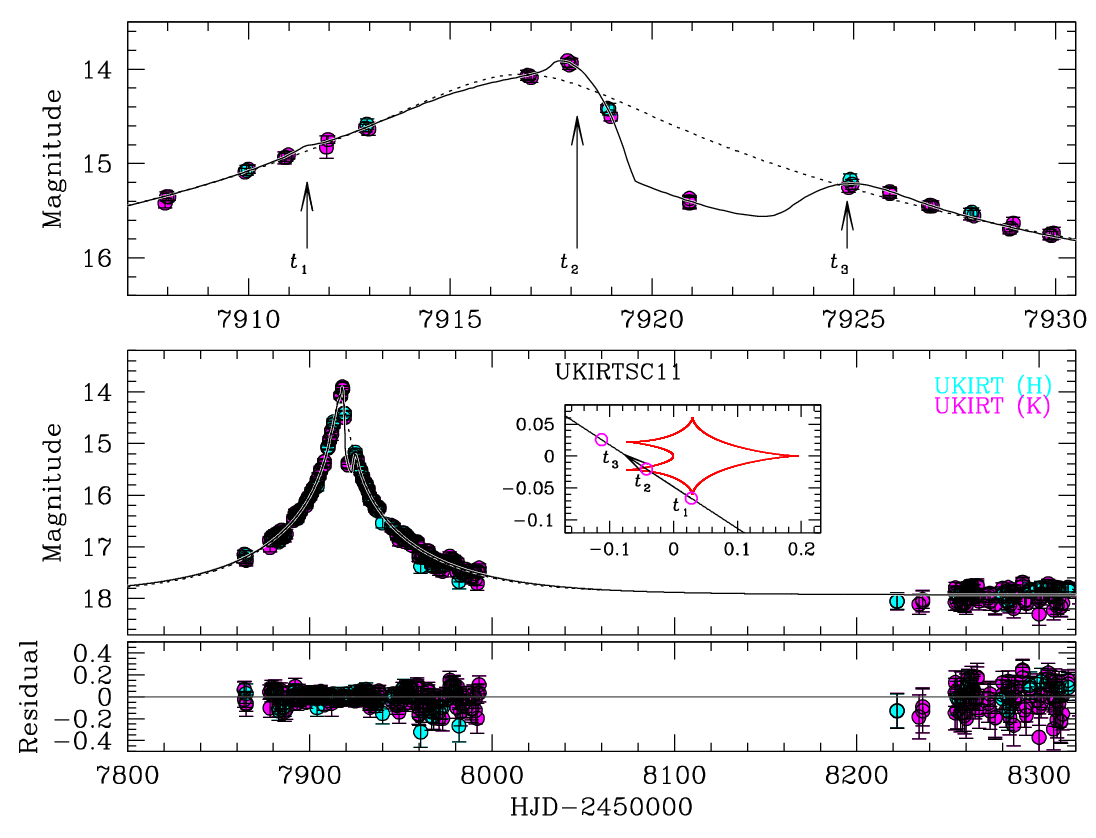}
\caption{
The lensing light curve of UKIRT11.  In the inset of the lower panel, the three empty circles on 
the source trajectory, labeled $t_1$, $t_2$, and $t_3$, represent the source positions at the times 
of consecutive approaches to the caustic cusps. The size of each circle is scaled to the size of 
the source. The times corresponding to these positions are indicated in the light curve by arrows.  
The dotted curve in the top panel represents a 1L1S model derived by excluding the data points 
around the anomaly.  
} 
\label{fig:ten}
\end{figure}

The lensing event UKIRT11, with its source situated very close to the Galactic center at Galactic 
coordinates $(l, b) = (-0^\circ\hskip-2pt .1218, -0^\circ\hskip-2pt .3291)$, was exclusively 
observed by the UKIRT survey during the 2017 season.  Figure~\ref{fig:ten} presents the lensing 
light curve for the event.  The source of this lensing event experienced a substantial 
magnification, with its peak brightness reaching roughly 4 magnitudes above the baseline level. 
Such a high-magnification event is especially sensitive to perturbations from a planetary companion 
to the lens, as pointed out by \citet{Griest1998}. The light curve indeed showed a short-term anomaly 
near the event's peak.  This anomaly exhibited both positive and negative deviations from the 1L1S 
model. The primary portion of the anomaly showed a negative deviation, flanked by weaker positive 
deviations on both sides of the main anomaly.

UKIRT11 is the only anomalous UKIRT lensing event that was previously analyzed in detail 
by \citet{Shvartzvald2018}, who reported that the companion in the binary lens system as a 
planetary object. Our modeling, utilizing the 2L1S configuration, independently confirmed 
that the anomaly arises from a planetary companion. The full set of lensing parameters 
for this solution is provided in Table~\ref{table:four}. The derived planetary parameters, 
$(s, q) \sim (1.01, 1.9 \times 10^{-3})$, are in close agreement with the values $(s, q) 
\sim (1.03, 1.5 \times 10^{-3})$ reported by \citet{Shvartzvald2018}.

The configuration of the lens system is depicted in the inset of the lower panel. The planet 
induces a resonant caustic, and the source passed close to this caustic multiple times. The 
source first approached the lower tip of the caustic at $t_1 = 7911.45$, then passed the 
lower-left cusp at $t_2 = 7918.15$, and later approached the upper-left cusp again at $t_3 = 
7918.15$. The source positions at these individual cusp approaches are marked with empty 
circles labeled $t_1$, $t_2$, and $t_3$, with arrows indicating their corresponding positions 
in the light curve. The size of the circles is scaled to represent the size of the source 
relative to the caustic. The major dip anomaly occurred between $t_2$ and $t_3$. The bump 
at $t_1$, caused by the source's initial approach to the cusp, is relatively weak due to 
the combined effects of a faint cusp and significant finite-source effects.

\begin{table*}[t]
\small
\caption{Best-fit parameters of UKIRT14, UKIRT15, UKIRT16, and UKIRT17.  \label{table:five}}
\begin{tabular}{lcccccc}
\hline\hline
\multicolumn{1}{c}{Parameter}    &
\multicolumn{1}{c}{UKIRT14}      &
\multicolumn{1}{c}{UKIRT15}      &
\multicolumn{1}{c}{UKIRT16}      &
\multicolumn{2}{c}{UKIRT17}      \\
\multicolumn{1}{c}{}             &
\multicolumn{1}{c}{}             &
\multicolumn{1}{c}{}             &
\multicolumn{1}{c}{}             &
\multicolumn{1}{c}{2L1S}         &
\multicolumn{1}{c}{1L2S}         \\
\hline 
 $\chi^2$             & 10629.2                &  9666.4                 &  501.2                &  379.2               &  378.0                \\
 $t_0$ (HJD$^\prime$) & $7990.33 \pm 0.25   $  &  $7977.625 \pm 0.025$   &  $8283.19 \pm 0.31 $  &  $8283.55 \pm 0.43$  &  $8265.79 \pm 0.27  $ \\
 $u_0$                & $0.60 \pm 3.98      $  &  $-1.307 \pm 0.038  $   &  $0.1176 \pm 0.0074$  &  $0.160 \pm 0.016 $  &  $0.114 \pm 0.021   $ \\
 $\te$ (days)         & $35.746 \pm 0.062   $  &  $65.69 \pm 0.42    $   &  $68.63 \pm 3.86   $  &  $168.33 \pm 18.62$  &  $90.32 \pm 11.17   $ \\
 $s$                  & $1.77344 \pm 0.00040$  &  $2.591 \pm 0.012   $   &  $0.67 \pm 0.02    $  &  $3.569 \pm 0.065 $  &  --                   \\
 $q$                  & $1.056 \pm 0.015    $  &  $1.193 \pm 0.036   $   &  $0.691 \pm 0.057  $  &  $3.384 \pm 1.063 $  &  --                   \\
 $\alpha$ (rad)       & $2.6275 \pm 0.0028  $  &  $5.8339 \pm 0.0042 $   &  $3.494 \pm 0.010  $  &  $5.365 \pm 0.012 $  &  --                   \\
 $\rho$ ($10^{-3}$)   & $2.980 \pm 0.024    $  &  $2.120 \pm 0.032   $   &  $2.33 \pm 0.33    $  &   --                 &  --                   \\
 $\pien$              & $-1.208 \pm 0.049   $  &   --                    &   --                  &   --                 &  --                   \\
 $\piee$              & $0.133 \pm 0.036    $  &   --                    &   --                  &   --                 &  --                   \\
 $ds/dt$              & $0.134 \pm 0.132    $  &   --                    &   --                  &   --                 &  --                   \\
 $d\alpha/dt$         & $-1.325 \pm 0.041   $  &   --                    &   --                  &   --                 &  --                   \\
 $t_{0,2}$            &  --                    &   --                    &   --                  &   --                 &  $8298.29 \pm 1.58  $ \\
 $u_{0,2}$            &  --                    &   --                    &   --                  &   --                 &  $0.272 \pm 0.078   $ \\
 $q_F$                &  --                    &   --                    &   --                  &   --                 &  $1.29 \pm 0.48     $ \\
\hline
\end{tabular}
\end{table*}

\begin{figure}[t]
\includegraphics[width=\columnwidth]{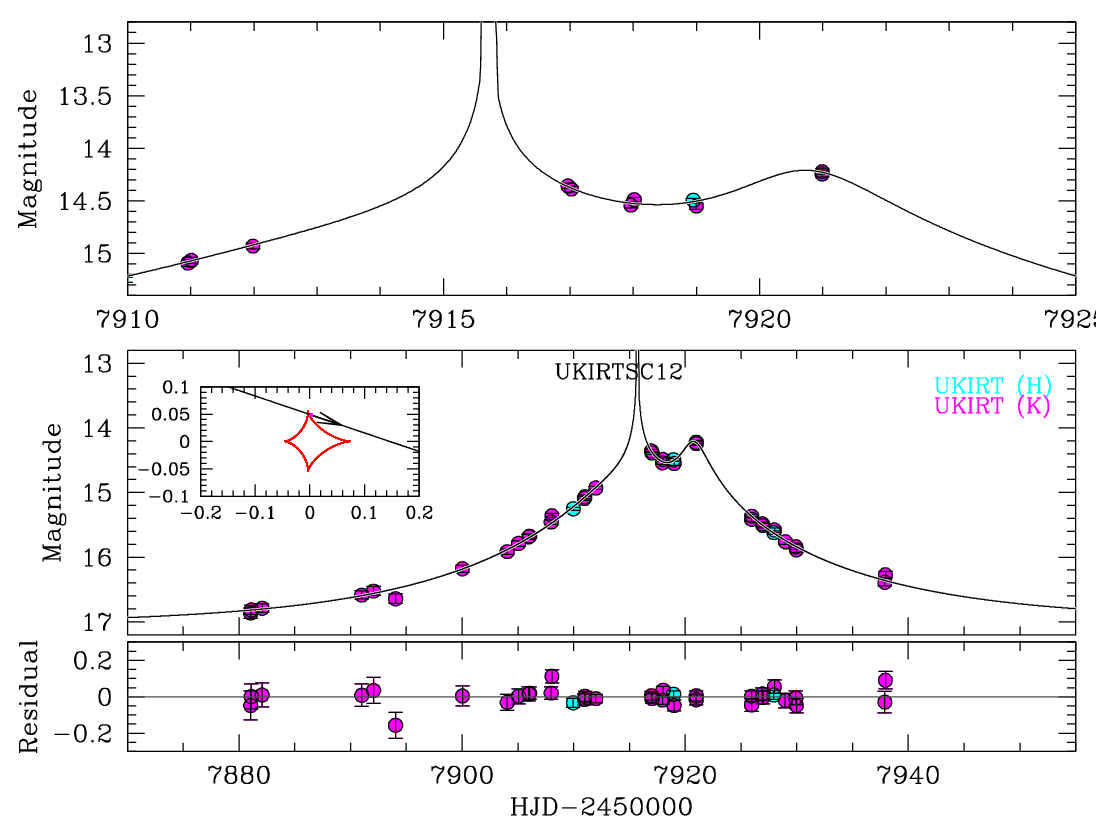}
\caption{
The lensing light curve of UKIRT12. 
}
\label{fig:eleven}
\end{figure}

\subsection{UKIRT12} \label{sec:four-ten}

The lensing event UKIRT12, located at Galactic coordinates $(l, b) = (1^\circ\hskip-2pt .243, 
-0^\circ\hskip-2pt .1533)$, was observed exclusively by the UKIRT survey during the 2017 
season, as it lies outside the coverage of the other surveys. Figure~\ref{fig:eleven} 
shows the lensing light curve of the event, which reached a moderately high magnification 
at its peak, with the brightness increasing by approximately 3.2 magnitudes above the 
baseline. Although observations were interrupted by unfavorable weather over a four-day 
period from $\hjd^\prime = 7913$ to 7916, the anomaly feature near the peak of the light 
curve is clearly discernible.

Modeling the light curve under a 2L1S configuration yielded a solution that well describe 
the anomaly feature. Unlike the event UKIRT11, the anomaly was turned out to be caused by 
a binary companion. The binary parameters of the solutions are $(s, q)\sim (3.7, 2.4)$, 
indicating that the companion has a mass greater than the lens component that the source 
approached and it is positioned at a separation substantially greater than the Einstein 
radius. Table~\ref{table:four} provides the complete lensing parameters of the solution.

The inset in the lower panel of Figure~\ref{fig:eleven} depicts the lens system configuration. 
The more massive binary lens companion generates a small diamond-shaped caustic near the 
position of the less massive lens component. The anomaly was caused by the source's diagonal 
passage through the vicinity of the caustic, first crossing the tip of the upper cusp and 
then approaching the on-axis cusp. The crossing of the first cusp produced a pronounced 
peak centered at around $\hjd^\prime = 7915.8$, but this feature was not observed due to 
a gap in coverage.

\begin{figure}[t]
\includegraphics[width=\columnwidth]{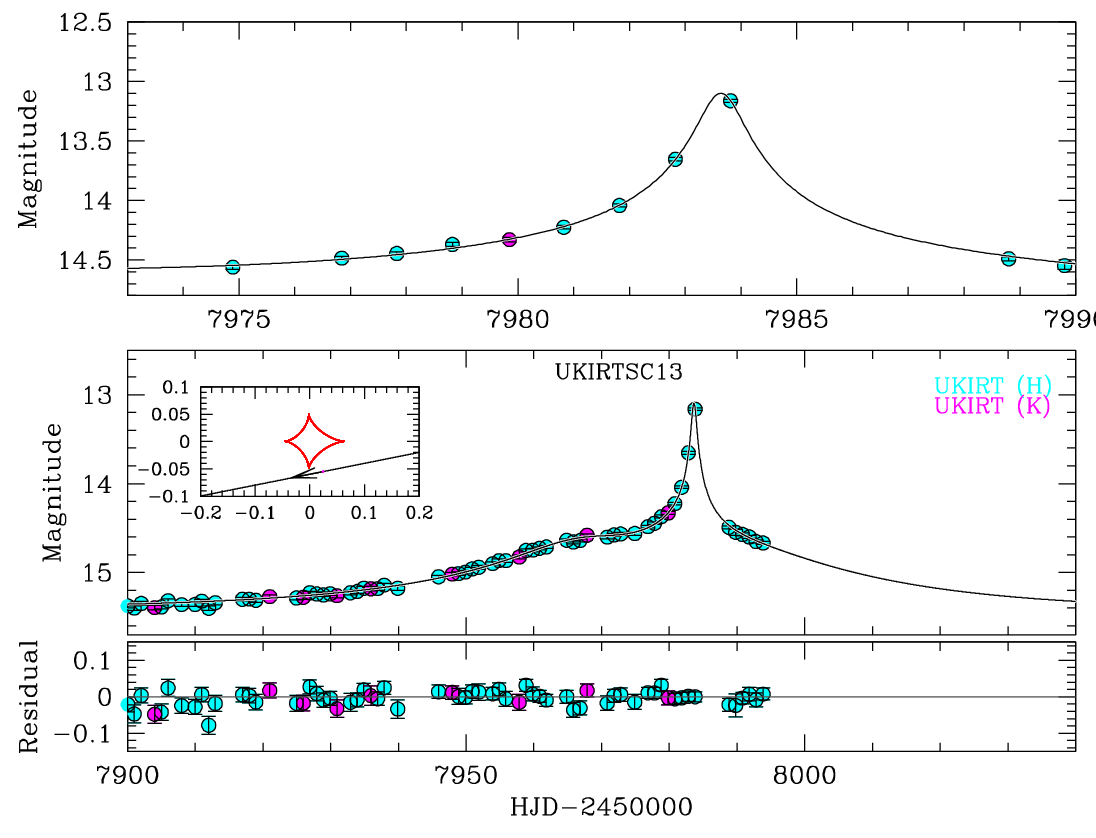}
\caption{
The lensing light curve of UKIRT13. 
}
\label{fig:twelve}
\end{figure}

\subsection{UKIRT13} \label{sec:four-eleven}

The lensing event UKIRT13, with its Galactic coordinates $(l, b)=(1^\circ\hskip-2pt .9771, 
-1^\circ\hskip-2pt .0028)$, was observed exclusively by the UKIRT survey conducted in the 
2017 season Figure~\ref{fig:twelve} shows the lensing light curve of the event. It exhibits 
a prominent anomaly centered at  $\hjd^\prime = 7993$, featuring strong positive deviations 
that exceed the baseline 1L1S model by about 1.7 magnitudes. The 2017 season ended at 
$\hjd^\prime = 7993$, and the light curve beyond this point was not covered.

The anomaly is well explained by a 2L1S model with binary lens parameters $(s, q) \sim 
(3.5, 5.9)$. The complete set of lensing parameters is provided in Table~\ref{table:four}. 
The model curve corresponding to this solution is overlaid on the data points in 
Figure~\ref{fig:twelve}. The inset in the lower panel illustrates the lens system configuration. 
This event is similar to UKIRT12, as the anomaly resulted from the source approaching the 
caustic induced by a wide-separation binary lens companion, with the companion having a 
greater mass than the lens component approached by the source. The event has a relatively 
long time scale, $\te \sim 120$~days. However, determining higher-order lensing parameters 
was challenging due to incomplete coverage of the light curve during the latter part of the 
event.

\begin{figure}[t]
\includegraphics[width=\columnwidth]{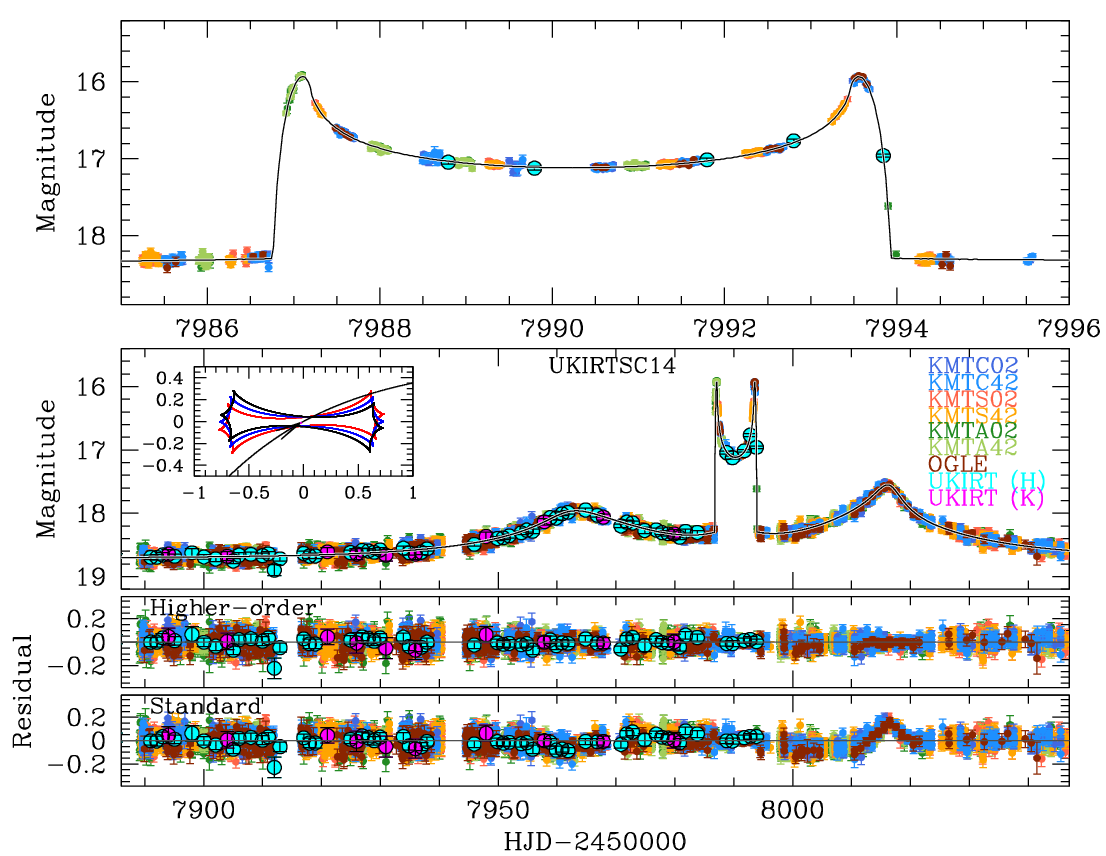}
\caption{
The lensing light curve of UKIRT14.  In the inset of the lower panel showing the lens-system 
configuration, the closed curves, marked in red, blue, and black, represent the caustics at 
$\hjd^\prime = 7960$, 7990, and 8017, respectively. The bottom two panels display the 
residuals from the models obtained using two approaches, one that considers higher-order 
effects and the other that does not.  
} \label{fig:thirteen}
\end{figure}

\subsection{UKIRT14} \label{sec:four-twelve}

The source of the lensing event UKIRT14, located at Galactic coordinates $(l, b)=(1^\circ
\hskip-2pt .4794, -1^\circ\hskip-2pt .5082)$, lies within the coverage of the three optical 
surveys conducted by the OGLE, KMTNet, and MOA groups conducted during the 2017 season. The 
event references designated by the individual surceys are OGLE-2017-BLG-1323, KMT-2017-BLG-0213, 
and MOA-2017-BLG-431.  Figure~\ref{fig:thirteen} presents 
the lensing light curve of the event, which is constructed from the combined data sets, 
excluding the MOA data due to relatively large photometric uncertainties. The light curve 
displays several distinct features, including caustic-crossing spikes at $\hjd^\prime = 
7987.0$ and 7981.5, as well as two bumps linked to cusp approaches at $\hjd^\prime = 7962$ 
and 8017.  The final bump was not captured by the UKIRT survey because the 2017 season ended 
before it could be observed, but it was densely covered by the data from the other surveys.

Modeling the light curve reveals that the anomalies are well explained by a binary lens. 
The binary lens parameters are $(s, q) = (1.77, 1.06)$, indicating that the lens consists 
of approximately equal-mass components with a projected separation larger than the Einstein 
radius. It was found that incorporating higher-order effects is essential to accurately 
describe the anomaly features. This is evident from the comparison of residuals, shown in 
the two bottom panels of Figure~\ref{fig:thirteen}, which were derived from two models -- 
one accounting for higher-order effects and the other not.  However, distinguishing between 
microlens-parallax and lens-orbital effects proved difficult, as both introduce similar 
deviations in the light curves. This overlap complicates the precise determination of the 
higher-order parameters.  The lensing parameters for one model are presented in 
Table~\ref{table:five}, but it is important to note that multiple local solutions exist, 
all providing similarly good fits to the data.

\begin{figure}[t]
\includegraphics[width=\columnwidth]{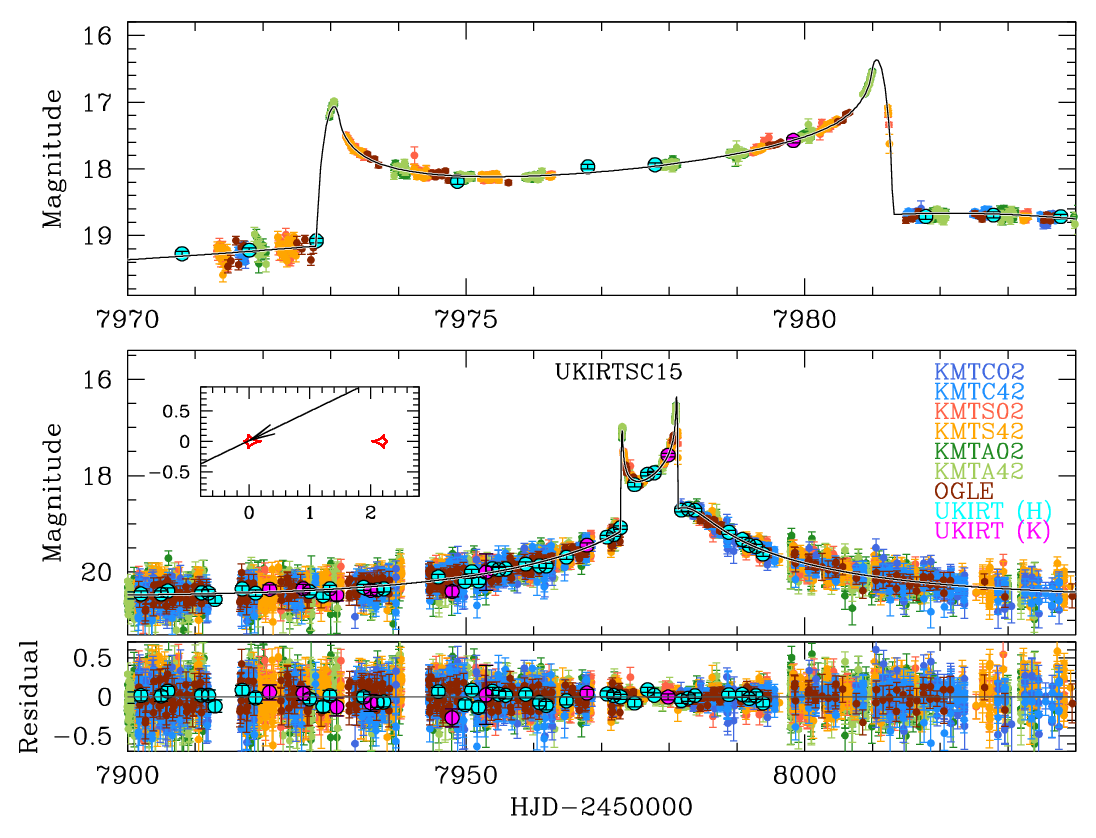}
\caption{
The lensing light curve of UKIRT15.
}
\label{fig:fourteen}
\end{figure}

The corresponding lens system configuration is illustrated in the inset of the lower panel. 
The caustic forms a resonant structure stretched along the binary axis. The source passed 
diagonally through the caustic, producing distinct caustic spikes. The two bumps, occurring 
before and after the caustic spikes, were caused by the source approaching the lower-left 
and upper-right cusps of the caustic, respectively.

\subsection{UKIRT15} \label{sec:four-thirteen}

The lensing event UKIRT15 was observed during the 2017 season by the UKIRT survey as well 
as three additional optical surveys: OGLE, KMTNet, and MOA. For this analysis, the MOA 
data set was excluded due to its relatively large uncertainties. The source position is 
within KMTNet's primary fields BLG02 and BLG43, toward which the event was monitored with 
a 0.25-hour cadence, resulting in dense light curve coverage. Figure~\ref{fig:fourteen} 
presents the lensing light curve of UKIRT15, constructed from the combined data sets. The 
light curve displays clear caustic-crossing features, with spikes at $\hjd^\prime = 7973.0$ 
and 7981.1. Both caustic spikes were well-covered by the combined KMTA and KMTS data sets. 
The portions of the light curve not captured by the UKIRT data after $\hjd^\prime = 7993$ 
were densely covered by data from the other surveys.

A 2L1S modeling of the light curve provides a unique solution characterized by the binary 
parameters $(s, q) \sim (2.6, 1.2)$ and an event time scale of $\te \sim 66$ days. 
Table~\ref{table:five} presents the full set of lensing parameters for the solution.
The lens system configuration, illustrated in the inset of the lower panel in 
Figure~\ref{fig:fourteen}, reveals that the binary lens creates two distinct sets of caustics, 
each located near one of the individual lens components. The source passed through the caustic 
associated with the lower-mass lens component, leading to the observed caustic spikes in the 
light curve.

\subsection{UKIRT16} \label{sec:four-fourteen}

\begin{figure}[t]
\includegraphics[width=\columnwidth]{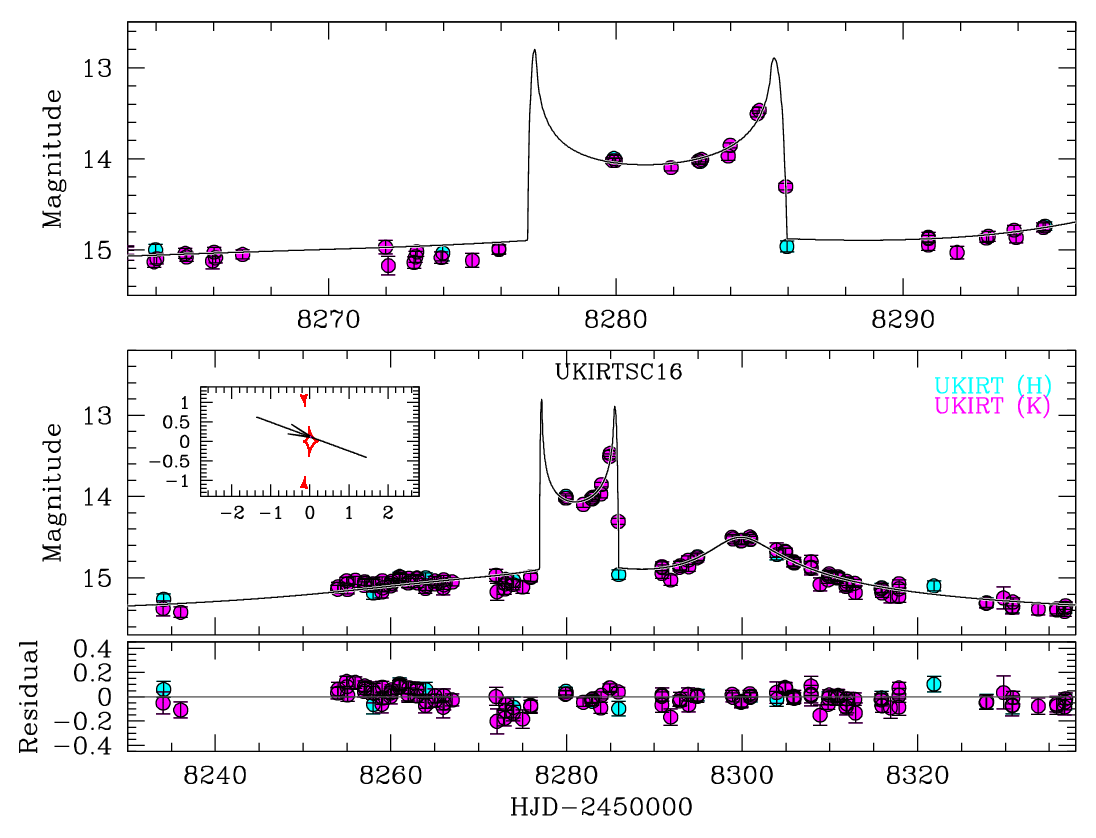}
\caption{
The lensing light curve of UKIRT16.
}
\label{fig:fifteen}
\end{figure}

UKIRT16 was the first anomalous lensing event reported from the UKIRT survey during the 2018
season. The source of the event, located near the Galactic center with Galactic coordinates 
$(l, b) = (0^\circ\hskip-2pt .5866, 0^\circ\hskip-2pt .4287)$, resides in a region not covered 
by other surveys. Figure~\ref{fig:fifteen} presents the lensing light curve of UKIRT16, which 
exhibits caustic-crossing feature at $\hjd^\prime =  8285.7$, followed by a bump around 
$\hjd^\prime = 8300$. While not captured in the data, an additional caustic spike is inferred 
around $\hjd^\prime = 8276.8$ from the extension of the U-shaped profile in the region between 
the observed caustic spikes.

\begin{table*}[t]
\small
\caption{Best-fit parameters of UKIRT18, UKIRT19, UKIRT20, UKIRT21, and UKIRT22.  \label{table:six}}
\begin{tabular}{lcccccc}
\hline\hline
\multicolumn{1}{c}{Parameter}    &
\multicolumn{1}{c}{UKIRT18}      &
\multicolumn{1}{c}{UKIRT19}      &
\multicolumn{1}{c}{UKIRT20}      &
\multicolumn{1}{c}{UKIRT21}      &
\multicolumn{2}{c}{UKIRT22}      \\
\multicolumn{1}{c}{}             &
\multicolumn{1}{c}{}             &
\multicolumn{1}{c}{}             &
\multicolumn{1}{c}{}             &
\multicolumn{1}{c}{}             &
\multicolumn{1}{c}{Close}        &
\multicolumn{1}{c}{Wide}         \\
\hline
 $\chi^2$                 &  10617.9               &  10687.2               &  5650.6              &  598.5                  &  621.8                 &  609.8                \\ 
 $t_0$ (HJD$^\prime$)     &  $8257.756 \pm 0.058$  &  $8285.770 \pm 0.022$  &  $8266.86 \pm 0.47$  &  $8621.474 \pm 0.109$   &  $8664.786 \pm 0.478$  &  $8665.801 \pm 0.497$ \\
 $u_0$                    &  $0.0280 \pm 0.0036 $  &  $0.4530 \pm 0.0036 $  &  $0.149 \pm 0.015 $  &  $0.0876 \pm 0.0059 $   &  $0.563 \pm 0.012   $  &  $0.664 \pm 0.019   $ \\
 $\te$ (days)             &  $38.99 \pm 0.53    $  &  $8.930 \pm 0.034   $  &  $137.95 \pm 11.07$  &  $17.02 \pm 0.85    $   &  $38.18 \pm 0.82    $  &  $33.21 \pm 0.73    $ \\
 $s$                      &  $0.8815 \pm 0.0061 $  &  $0.6274 \pm 0.0012 $  &  $0.50 \pm 0.017  $  &  $0.574 \pm 0.019   $   &  $0.8701 \pm 0.0071 $  &  $1.5005 \pm 0.0064 $ \\
 $q$                      &  $1.444 \pm 0.023   $  &  $1.225 \pm 0.045   $  &  $0.442 \pm 0.048 $  &  $0.327 \pm 0.054   $   &  $0.311 \pm 0.034   $  &  $0.107 \pm 0.017   $ \\
 $\alpha$ (rad)           &  $2.190 \pm 0.010   $  &  $2.060 \pm 0.013   $  &  $3.313 \pm 0.020 $  &  $5.760 \pm 0.025   $   &  $2.379 \pm 0.036   $  &  $4.271 \pm 0.023   $ \\
 $\rho$ ($10^{-3}$)       &  $4.86 \pm 0.23     $  &  $8.31 \pm 0.13     $  &   --                 &  $30.11 \pm 3.47    $   &  $11.66 \pm 2.39    $  &  $20.22 \pm 2.00    $ \\
 $t_{0,2}$ (HJD$^\prime$) &   --                   &   --                   &   --                 &  $8635.488 \pm 0.113$   &   --                   &                       \\
 $u_{0,2}$                &   --                   &   --                   &   --                 &  $-0.328 \pm 0.042  $   &   --                   &                       \\
 $\rho_2$ ($10^{-3}$)     &   --                   &   --                   &   --                 &  $227.14 \pm 19.69  $   &   --                   &                       \\
 $q_F$                    &   --                   &   --                   &   --                 &  $1.93 \pm 0.23     $   &   --                   &                       \\
\hline
\end{tabular}
\end{table*}

Considering the caustic-related anomaly features, we performed 2L1S modeling of the light 
curve. The modeling yields a solution with binary parameters $(s, q) \sim (0.67, 0.69)$ and 
an event time scale of $\te \sim 69$ days. The complete set of lensing parameters is listed 
in Table~\ref{table:five}. The normalized source radius is derived from the analysis of the 
caustic exit that was partially covered by the data.  As shown in the inset of the lower 
panel, the lens system configuration reveals that the binary lens produced three sets of 
caustics, with one near the barycenter of the binary lens and two located farther away. 
The caustic features arise from the source passing through the central caustic.

\subsection{UKIRT17} \label{sec:four-fifteen}

\begin{figure}[t]
\includegraphics[width=\columnwidth]{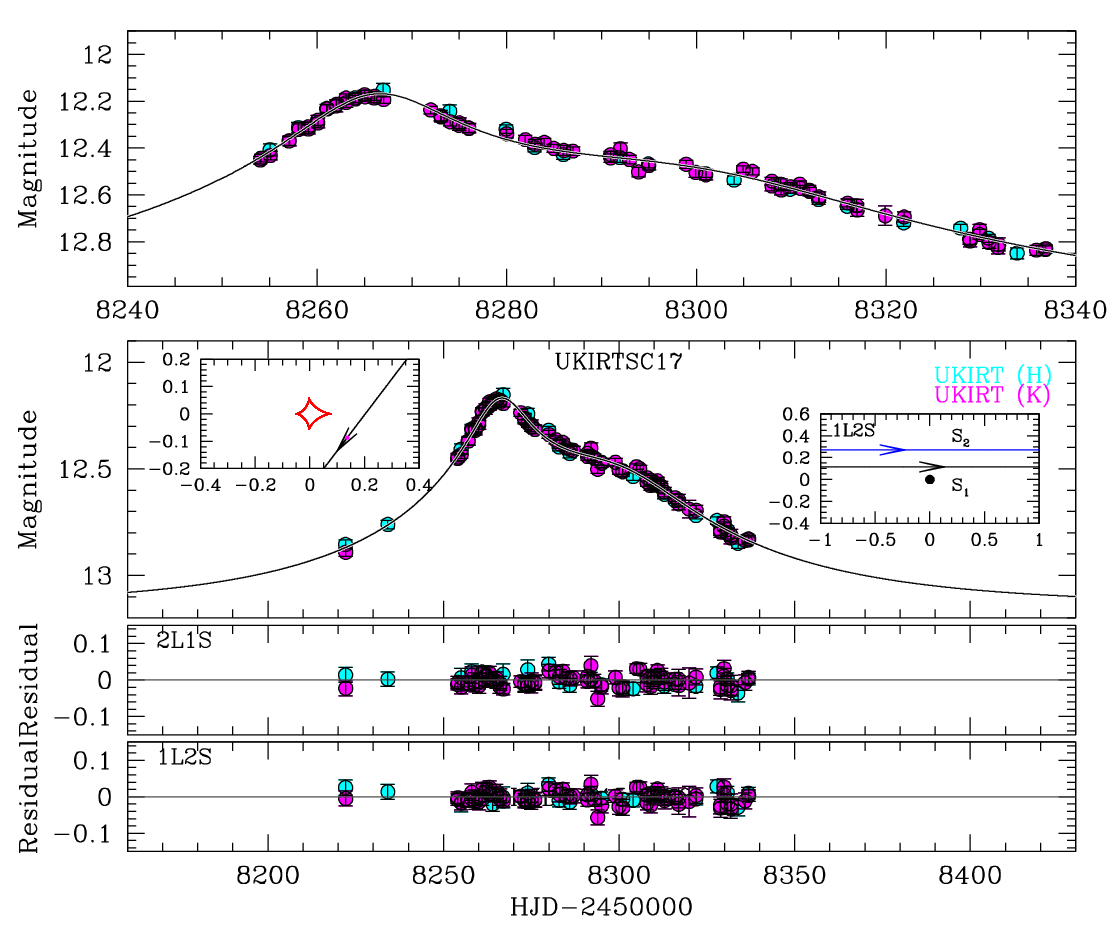}
\caption{
The lensing light curve of UKIRT17.  The model curves for the 2L1S and 1L2S models are 
superimposed on the data points, appearing indistinguishable within the thickness of the 
lines.
}
\label{fig:sixteen}
\end{figure}

The source of the lensing event UKIRT17, with Galactic coordinates $(l, b) \sim (1^\circ\hskip-2pt 
.5664, -0^\circ\hskip-2pt .4300)$, lies in the region exclusively covered by the UKIRT survey 
during the 2018 season. The lensing light curve of the event is presented in Figure~\ref{fig:sixteen}. 
It does not display a discontinuous anomaly but rather exhibits a smooth deviation from the 1L1S 
model.

Considering the characteristics of the anomaly, we carried out modeling using both the 
2L1S and 1L2S models. The results show that the anomaly is explained almost equally well 
by both models, with $\chi^2=379.2$ for the 2L1S model and $\chi^2=378.0$ for the 1L2S model.

The lensing parameters for both models are listed in Table~\ref{table:five}. The lens system 
configurations corresponding to the 2L1S and 1L2S solutions are illustrated in the left and 
right insets of the lower panel in Figure~\ref{fig:sixteen}, respectively.  Under the 2L1S 
solution, the lens is a binary characterized by $(s, q) \sim (2.6, 1.2)$, and the source 
passed through the outer region of a small caustic formed by the binary lens. In the 1L2S 
interpretation, the source is a binary consisting of two stars with a flux ratio of $q_F 
\sim 1.3$. The model curves for both solutions are superimposed on the data points, appearing 
indistinguishable within the thickness of the lines.

\begin{figure}[t]
\includegraphics[width=\columnwidth]{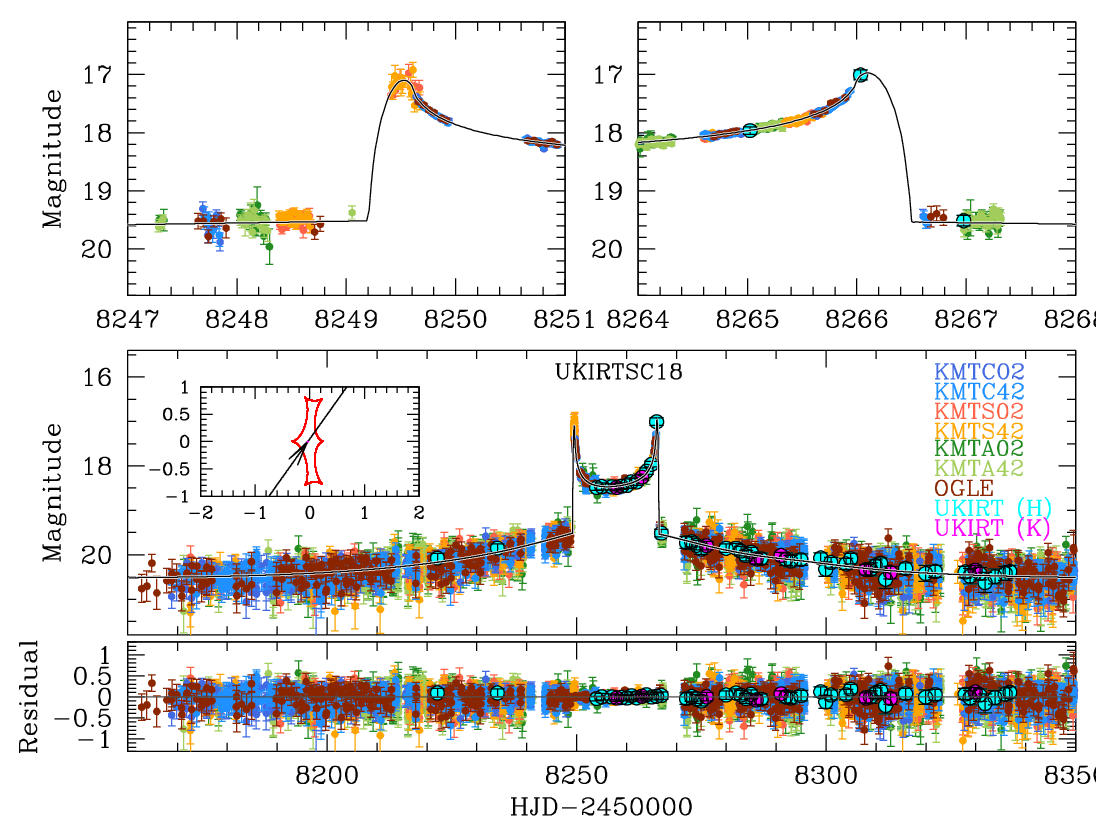}
\caption{
The lensing light curve of UKIRT18.  The two top panels show zoomed-in views of the sections 
during the source entrance and exit from the caustic.
}
\label{fig:seventeen}
\end{figure}

\subsection{UKIRT18} \label{sec:four-sixteen}

The lensing event UKIRT18, located at Galactic coordinates $(l, b) = (1^\circ\hskip-2pt .2397, 
-1^\circ\hskip-2pt .0092)$, was observed during the 2018 season by the UKIRT, OGLE, and KMTNet 
surveys. The OGLE group designated the event as OGLE-2018-BLG-0752.  Although the event was 
not initially detected in the KMTNet survey, and therefore no KMTNet ID was assigned, its 
photometric data was later retrieved through postseason photometry of the source.

Figure~\ref{fig:seventeen} displays the lensing light curve of UKIRT18, which was constructed 
by combining data from the UKIRT, OGLE, and KMTNet surveys. Although the observational cadence 
of the UKIRT survey was relatively low, the high cadences of the OGLE and KMTNet surveys 
allowed for a densely sampled light curve. This high-resolution data revealed intricate details 
of the light curve, including distinctive caustic-crossing features. The two spikes, centered 
at $\hjd^\prime \sim 8249.5$ and 8266.0, suggest the binary nature of the lens.

\begin{figure}[t]
\includegraphics[width=\columnwidth]{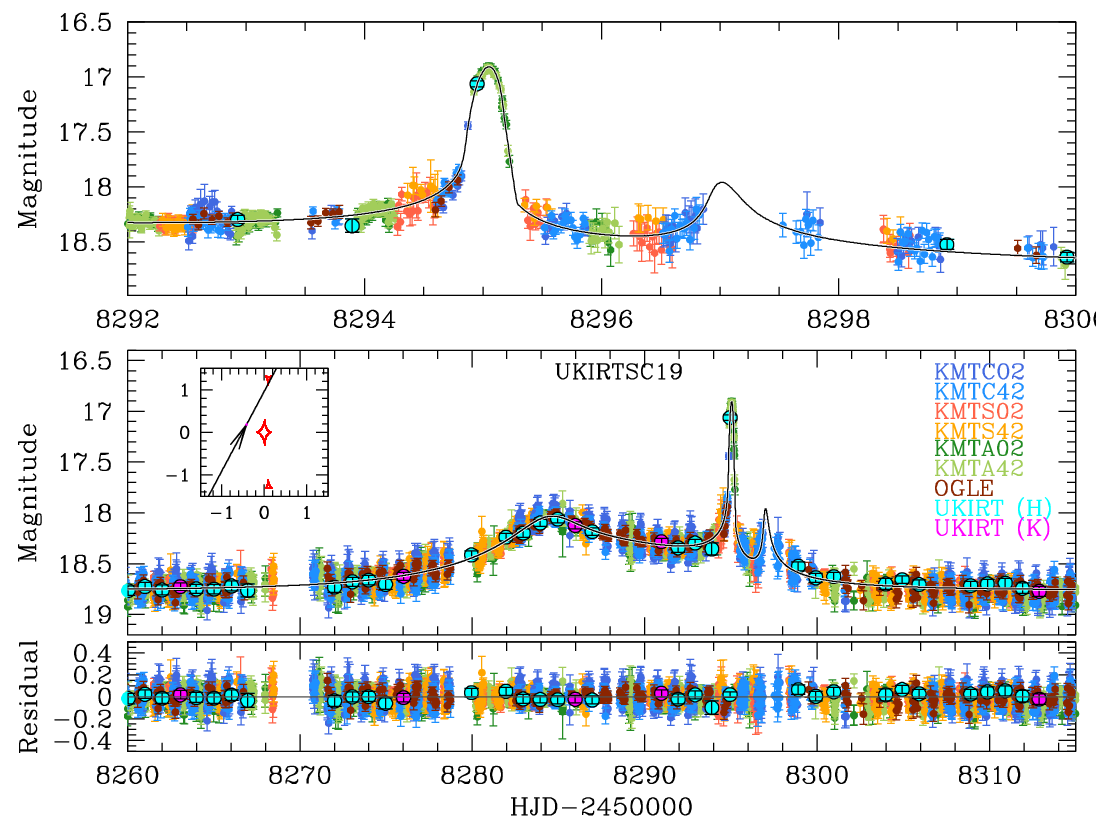}
\caption{
The lensing light curve of UKIRT19.  
}
\label{fig:eighteen}
\end{figure}

Modeling using the 2L1S configuration yields a unique solution with binary parameters $(s, q) 
\sim (0.88, 1.44)$ and an event time scale of $\te \sim 39$ days.  The complete set of lensing 
parameters for this solution is listed in Table~\ref{table:six}. The corresponding model curve, 
which accurately fits the data, is overlaid on the data points in Figure~\ref{fig:seventeen}. 
In the inset of the lower panel, both the caustic and the source trajectory are illustrated. 
The binary lens generates a resonant caustic that extends in a direction perpendicular to the 
binary axis. The source passes diagonally through this caustic, initially entering through 
the upper right fold and exiting via the lower left fold. The normalized source radius was 
determined by analyzing the data during the caustic crossing, which is well-resolved in the 
light curve.

\begin{figure}[t]
\includegraphics[width=\columnwidth]{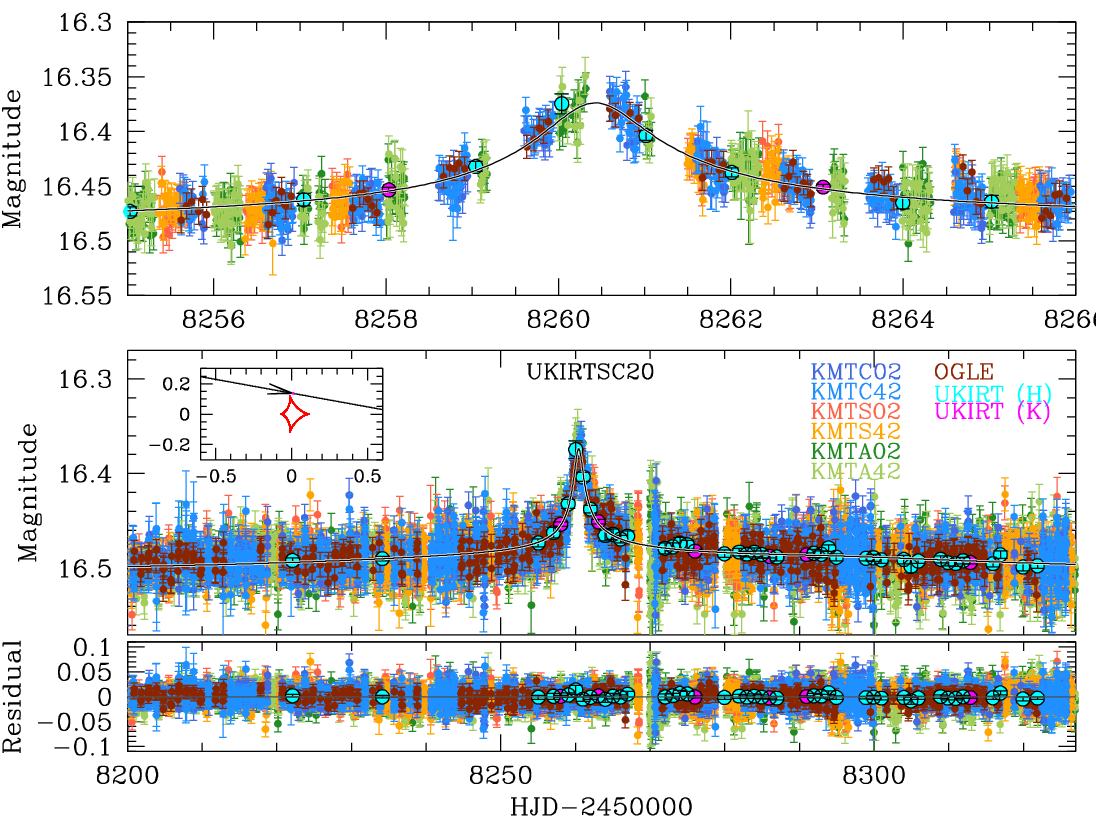}
\caption{
The lensing light curve of UKIRT20.  
}
\label{fig:nineteen}
\end{figure}

\subsection{UKIRT19} \label{sec:four-seventeen}

The lensing event UKIRT19, detected during the 2018 season, was also observed by the OGLE 
and KMTNet surveys, which designated the event as OGLE-2018-BLG-1055 and KMT-2018-BLG-2095, 
respectively. The light curve, constructed by combining data from the three surveys, is 
shown in Figure~\ref{fig:eighteen}. It exhibits a relatively complex pattern, characterized 
by two successive positive deviations occurring at $\hjd^\prime \sim 8295.0$ and 8297.1, 
both appearing on the descending side of the curve. Since the maximum deviation of the first 
bump exceeds 1.5 magnitudes, it is likely that the anomaly was caused by a caustic crossing.  
Another noteworthy characteristic is that, excluding these anomalous features, the light 
curve is asymmetric.

From the 2L1S modeling, we obtain a unique solution that successfully explains all the 
observed anomaly features. The binary lens parameters of this solution are $(s, q) \sim 
(0.63, 1.23)$. The estimated event time scale, $\te \sim 9$ days, is relatively short 
compared to other events. The complete set of lensing parameters is provided in 
Table~\ref{table:six}. The model curve and residuals from the fit are shown in 
Figure~\ref{fig:eighteen}. The lens system configuration, depicted in the inset of the 
lower panel, illustrates that the anomaly was caused by the source's approach to one of 
the two peripheral caustics formed by a binary lens with a separation smaller than the 
Einstein radius.  The two bumps in the anomaly occur because the source passed near the 
two cusps of the three-cusp peripheral caustic.

\subsection{UKIRT20} \label{sec:four-eighteen}

UKIRT20 is another lensing event commonly observed in the 2018 season by the three lensing
surveys of the UKIRT, OGLE, and KMTNet.  The ID references designated by the OGLE and KMTNet 
surveys are OGLE-2018-BLG-0856 and KMT-2018-BLG-2392, respectively.  Figure~\ref{fig:nineteen} 
shows the lensing light curve of the event, showing that the source flux exhibited a rapid 
rise and fall during $8250 \lesssim \hjd^\prime \lesssim 8260$, which were captured by 
the data from all three surveys. The source flux magnification due to lensing, excluding 
the anomaly, was small but persisted for a considerable duration.

The analysis reveals that the anomaly is well explained by a 2L1S model. The binary 
parameters for the solution are $(s, q) \sim (0.50, 0.44)$. As anticipated from the 
extended duration of the lensing magnification, the estimated event time scale, $\te 
\sim 138$ days, is relatively long compared to typical events. The complete set of 
lensing parameters is provided in Table~\ref{table:six}, and the model curve corresponding 
to the solution is overlaid on the data points in Figure~\ref{fig:nineteen}. The lens system 
configuration, shown in the inset of the lower panel, indicates that the binary lens produces 
a fold-fold caustic near the barycenter of the binary.  The anomaly was caused by the source 
approaching the upper cusp of the caustic.

\subsection{UKIRT21} \label{sec:four-nineteen}

\begin{figure}[t]
\includegraphics[width=\columnwidth]{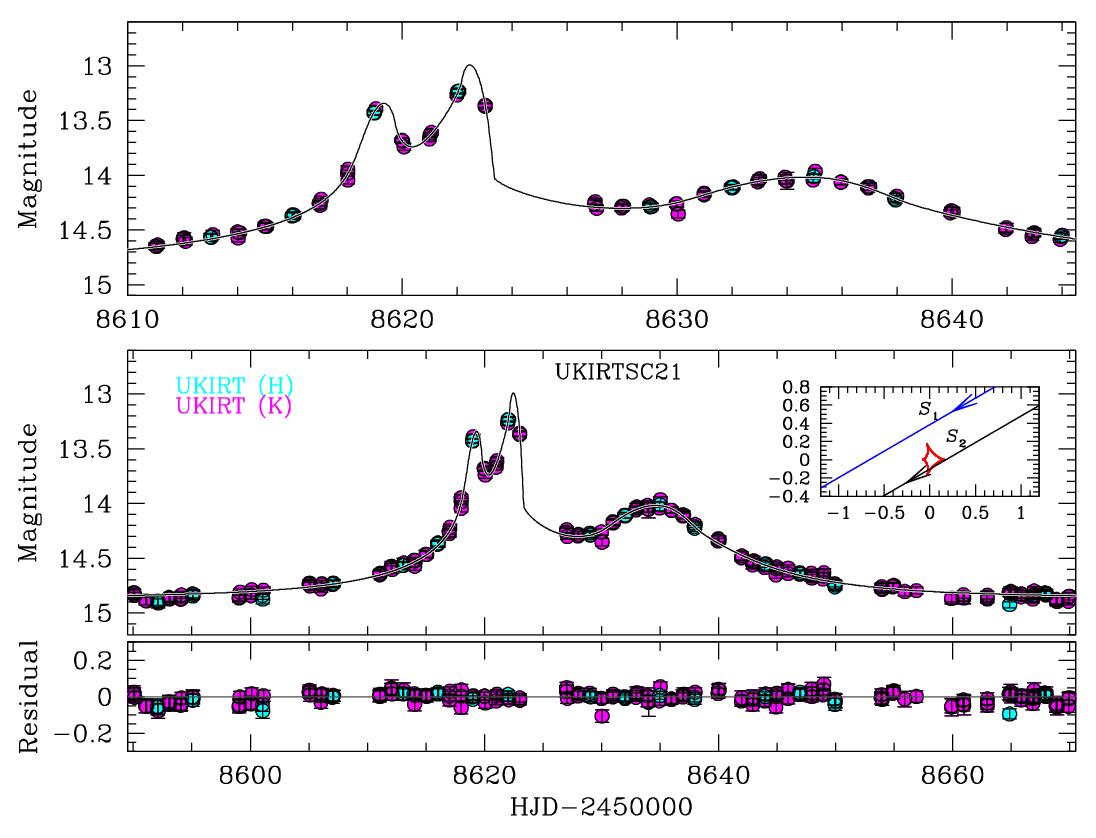}
\caption{
The lensing light curve of UKIRT21.  
}
\label{fig:twenty}
\end{figure}

UKIRT21 is the first anomalous lensing event detected during the 2019 season by the 
UKIRT survey. The source of the event is located near the Galactic center, with Galactic 
coordinates $(l, b) = (-1^\circ\hskip-2pt .0598, 0^\circ\hskip-2pt .3193)$, placing it 
outside the coverage of other optical lensing surveys. The lensing light curve of the 
event, shown in Figure~\ref{fig:twenty}, reveals multiple anomaly features. Between 
$\hjd^\prime \sim 8617$ and $\hjd^\prime \sim 8625$, the light curve displays 
characteristics indicative of caustic crossings. In addition to these features, the 
light curve also exhibits a bump centered at $\hjd^\prime \sim 8635$.

Considering the caustic-crossing features, we initially modeled the light curve using a 2L1S 
configuration. However, this approach did not fully capture all the anomalous features. To 
explore the possibility that the caustic anomaly was caused by a binary lens, while the bump 
feature resulted from additional components in the lens or source system, we performed a 
second 2L1S modeling, excluding the region around the bump. This model effectively described 
the caustic-crossing portion of the light curve. Next, we introduced an additional source 
component, resulting in a binary-lens and binary-source (2L2S) system. This modeling provided 
a solution that accurately accounted for all the anomalies in the light curve. A subsequent 
3L1S modeling did not yield a solution that matched the 2L2S model.

\begin{table*}[t]
\caption{Best-fit parameters of UKIRT23, UKIRT25, UKIRT26, and UKIRT27.  \label{table:seven}}
\begin{tabular}{lcccccc}
\hline\hline
\multicolumn{1}{c}{Parameter}    &
\multicolumn{1}{c}{UKIRT23}      &
\multicolumn{1}{c}{UKIRT25}      &
\multicolumn{1}{c}{UKIRT26}      &
\multicolumn{1}{c}{UKIRT27}      \\
\hline
 $\chi^2$                 &  656.6                &  483.9                &  10293.2                 &  8090.5                 \\     
 $t_0$ (HJD$^\prime$)     &  $8598.88 \pm 0.17 $  &  $8624.79 \pm 2.17$   &  $8670.9025 \pm 0.0092$  &  $8709.621 \pm 0.022$   \\
 $u_0$                    &  $0.1050 \pm 0.0017$  &  $0.590 \pm 0.022 $   &  $0.0606 \pm 0.0008   $  &  $0.2191 \pm 0.0013 $   \\
 $\te$ (days)             &  $59.27 \pm 0.54   $  &  $78.80 \pm 2.35  $   &  $8.427 \pm 0.016     $  &  $33.29 \pm 0.14    $   \\
 $s$                      &  $1.5542 \pm 0.0057$  &  $1.210 \pm 0.017 $   &  $1.3233 \pm 0.0015   $  &  $1.7303 \pm 0.0026 $   \\
 $q$                      &  $0.925 \pm 0.043  $  &  $1.72 \pm 0.32   $   &  $0.3966 \pm 0.0029   $  &  $1.092 \pm 0.011   $   \\
 $\alpha$ (rad)           &  $2.7891 \pm 0.0058$  &  $5.724 \pm 0.036 $   &  $-0.2459 \pm 0.0006  $  &  $5.1476 \pm 0.0017 $   \\
 $\rho$ ($10^{-3}$)       &  $5.97 \pm 1.27    $  &  $4.80 \pm 0.55   $   &  $16.001 \pm 0.079    $  &  $7.281 \pm 0.045   $   \\
\hline
\end{tabular}
\end{table*}

The model curve for the 2L2S solution is overlaid on the data points in the light curve, with 
the corresponding lensing parameters listed in Table~\ref{table:six}. According to this solution, 
the lens is a binary system with parameters $(s, q) \sim (0.57, 0.33)$, creating a small four-cusp 
caustic near the barycenter of the binary lens, as shown in the inset of the lower panel. The 
source is also binary, consisting of two stars with a flux ratio of $q_F \sim 1.9$. The inset 
in the lower panel of Figure~\ref{fig:twenty} illustrates the trajectories of the two source 
stars relative to the caustic. The caustic-related anomaly was caused by the fainter source 
star, which passed near the tip of the right-side on-axis cusp and then crossed the lower cusp. 
This sequence of caustic crossings, combined with significant finite-source effects, caused the 
anomaly to deviate from the typical caustic-crossing pattern. The bump feature, on the other 
hand, was produced by the approach of the brighter source star, which followed the fainter star 
and passed with a large impact parameter.

\begin{figure}[t]
\includegraphics[width=\columnwidth]{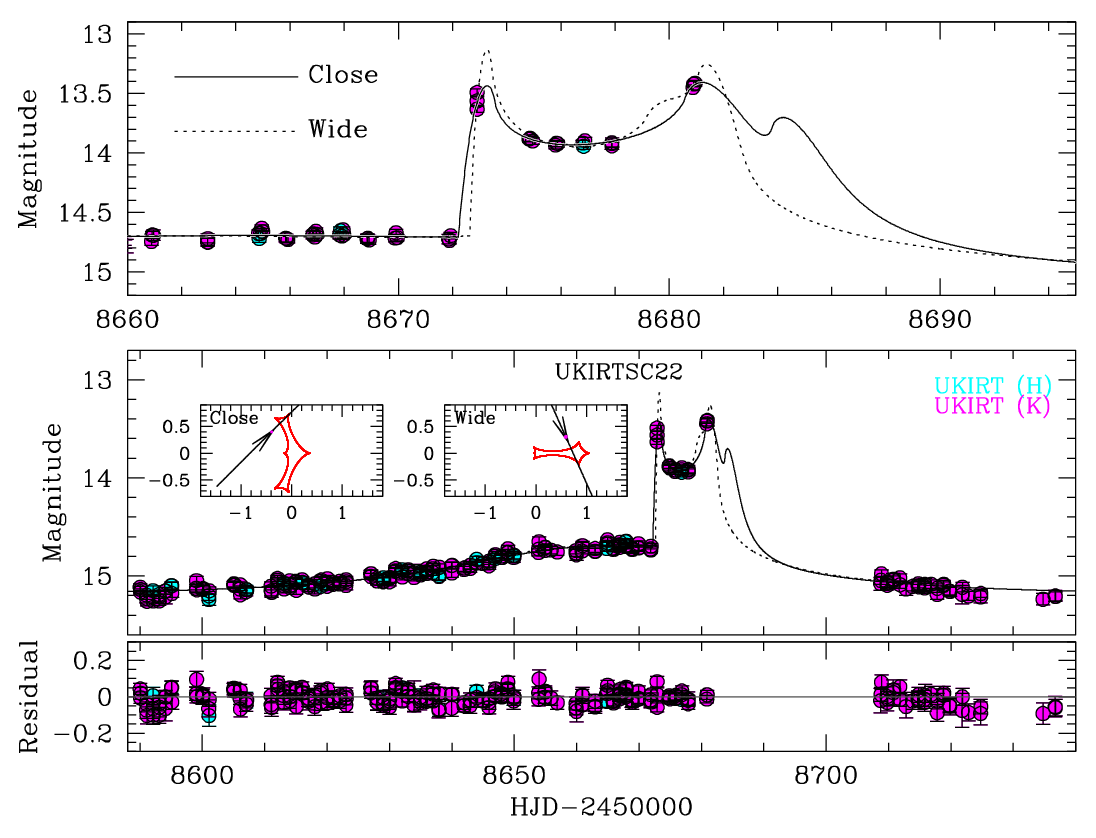}
\caption{
The lensing light curve of UKIRT22. The two insets of the lower panel show the lens system
configurations of the close and wide solutions.
}
\label{fig:twentyone}
\end{figure}

\subsection{UKIRT22} \label{sec:four-twenty}

The lensing event UKIRT22, which occurred during the 2019 season, was observed exclusively 
by the UKIRT survey, as its source is located near the Galactic center, with Galactic 
coordinates $(l, b) = (-0^\circ\hskip-2pt .8297, 0^\circ\hskip-2pt .2613)$, placing it in 
a region not covered by other surveys.  Figure~\ref{fig:twentyone} presents the lensing 
light curve of the event, which shows caustic-crossing features. One spike is clearly 
visible at  $\hjd^\prime \sim 8671.9$, while the other caustic spike is not resolved by 
the data. The anomaly occurred during a period of low lensing magnification.

Modeling the light curve using a 2L1S configuration results in a pair of degenerate solutions: 
a close solution with $(s, q) \sim (0.87, 0.31)$ and a wide solution with $(s, q) \sim (1.50, 
0.11)$. These solutions are referred to as "close" and "wide," depending on whether the binary 
separation is smaller or larger than the Einstein radius, respectively.  The model curves of 
the solutions are drawn over data points in Figure~\ref{fig:twentyone}.  The complete set of 
the lensing parameters are provided in Table~\ref{table:six}.  While the wide solution provides 
a slightly better fit, the difference in $\chi^2$ values, $\Delta \chi^2 \sim 11.0$, is minor.

The lens system configurations for the close and wide solutions are shown in the inset of the 
lower panel of Figure~\ref{fig:twentyone}. The caustics for both solutions exhibit a resonant 
structure, with the caustic of the close solution extending perpendicular to the binary lens 
axis, while that of the wide solution elongates along the axis. In both cases, the source 
crossed the edge of the caustic. After exiting, the source briefly crossed the tip of the 
caustic, resulting in a light curve profile that deviates from the typical caustic-crossing
 pattern.

\begin{figure}[t]
\includegraphics[width=\columnwidth]{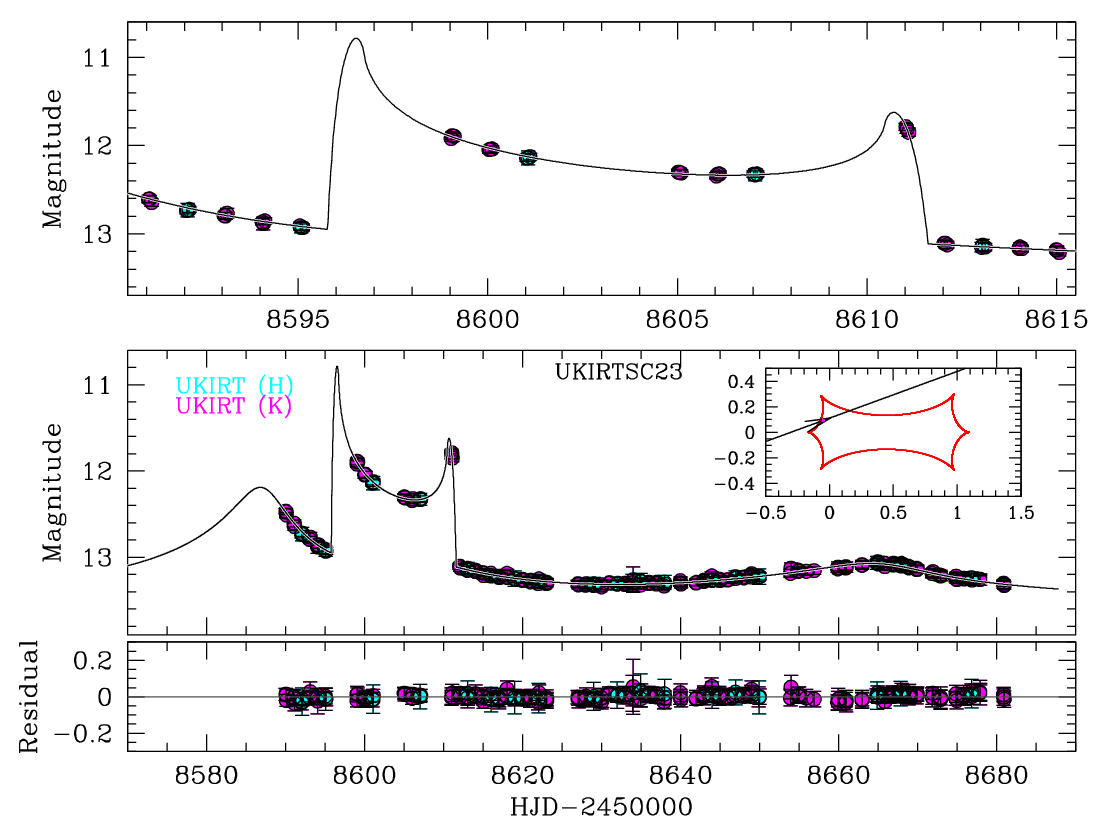}
\caption{
The lensing light curve of UKIRT23. 
}
\label{fig:twentytwo}
\end{figure}

\subsection{UKIRT23} \label{sec:four-twentyone}

The source of the event UKIRT23, located near the Galactic center at Galactic coordinates 
$(l, b) = (-0^\circ\hskip-2pt .3900, 0^\circ\hskip-2pt .5227)$, was observed exclusively 
by the UKIRT survey conducted in the 2019 season. Figure~\ref{fig:twentytwo} shows the 
lensing light curve of the event, which features prominent caustic-crossing signatures. 
The first unresolved caustic spike is estimated to have occurred around $\hjd^\prime \sim 
8696$, while the second spike is observed at $\hjd^\prime \sim 8611$.  In addition to these 
caustic-crossing features, the light curve shows two additional bumps, with a prominent one 
appearing before the first caustic spike and a weaker one centered near $\hjd^\prime \sim 
8666$.

Modeling performed under the 2L1S lens system configuration yielded a unique solution 
that accurately reproduces all features of the anomaly.  The binary parameters of the 
solution are $(s, q) \sim (1.55, 0.93)$, and the estimated time scale of the event is 
$\te \sim 59$ days. Table~\ref{table:seven} provides the complete set of lensing parameters. 
As illustrated in the lower panel of Figure~\ref{fig:twentytwo}, the lens system forms a 
resonant caustic extended along the binary lens axis. The source initially approached the 
left on-axis cusp, resulting in the first bump. It then entered the caustic by crossing 
the upper-left fold, producing the first caustic spike. Subsequently, the source exited 
the caustic by traversing the upper fold, which caused the second caustic spike. A final 
weak bump occurred as the source neared the upper-right cusp of the caustic.

\begin{figure}[t]
\includegraphics[width=\columnwidth]{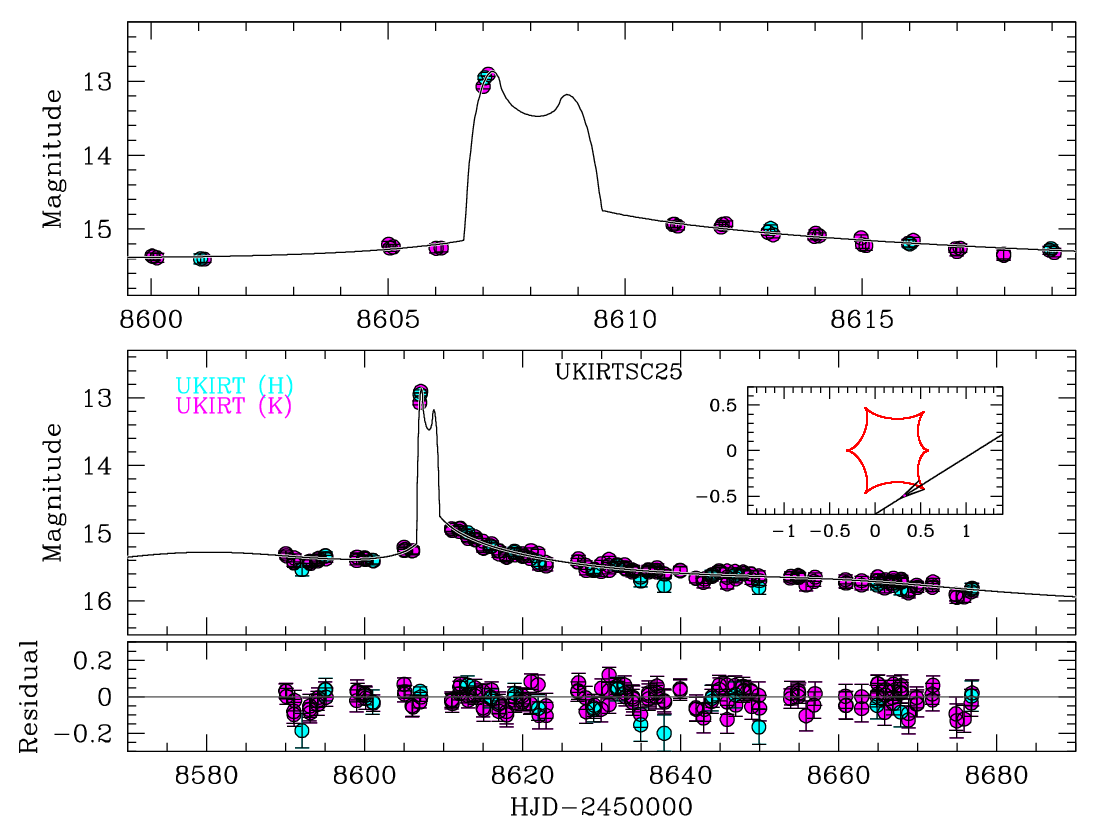}
\caption{
The lensing light curve of UKIRT25. 
}
\label{fig:twentythree}
\end{figure}

\subsection{UKIRT25} \label{sec:four-twentytwo}

The lensing event UKIRT25, observed exclusively by the UKIRT survey, was detected during 
the 2019 season.  Figure~\ref{fig:twentythree} presents the lensing light curve of the 
event, which shows a caustic-crossing feature around $\hjd^\prime \sim 8608$, with three 
data points capturing the first caustic spike.  Although the duration of the caustic-related 
feature is relatively short, about 5 days, the anomaly is unlikely to result from a planetary 
companion, as the overall light curve deviates from the symmetric 1L1S profile. The occurrence 
of the caustic-related anomaly at low magnification suggests that the source crossed a portion 
of the caustic located away from the barycenter of the binary lens.

Binary-lensing modeling has identified a solution that accurately accounts for the anomaly 
observed in the lensing light curve. The binary parameters corresponding to this solution 
are $(s, q) \sim (1.20, 1.72)$, with an event time scale of $\te \sim 79$ days. The complete 
set of lensing parameters associated with this model is detailed in Table~\ref{table:seven}.  
As illustrated in the inset of the lower panel, the lens system configuration demonstrates 
that the binary lens generates a resonant caustic characterized by six distinct cusps.  The 
source trajectory passed the tip of the lower-right cusp, leading to the observed caustic-crossing 
anomaly, which occurred at a relatively low magnification. Furthermore, the asymmetry observed 
in the light curve results from the influence of the large caustic, which distorts the overall 
magnification pattern.

\begin{figure}[t]
\includegraphics[width=\columnwidth]{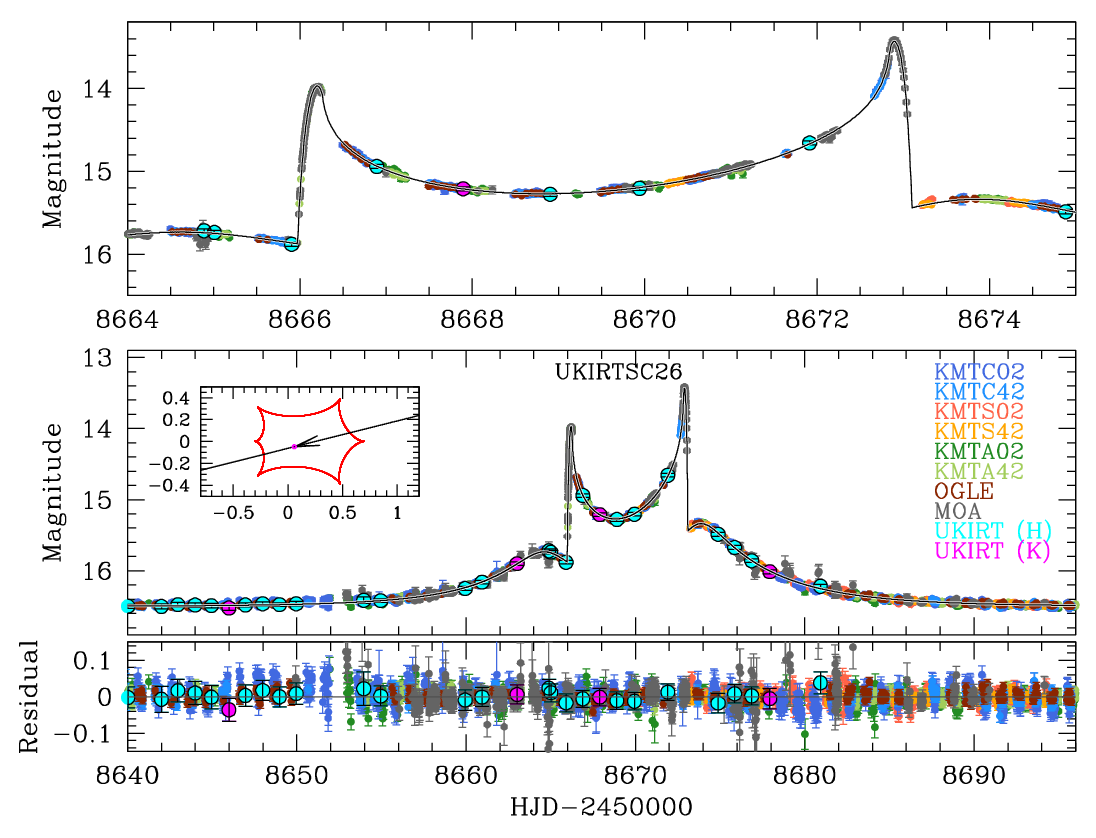}
\caption{
The lensing light curve of UKIRT26. 
}
\label{fig:twentyfour}
\end{figure}

\subsection{UKIRT26} \label{sec:four-twentythree}

The lensing event UKIRT26, which occurred in the 2019 season, was observed not only by the 
UKIRT survey but also by three optical lensing surveys conducted by the OGLE, KMTNet, and 
MOA groups. The source lies within the prime fields of these optical surveys, leading to a 
very dense coverage of the light curve, as shown in Figure~\ref{fig:twentyfour}. The light 
curve is characterized by well-resolved caustic-crossing features, with two distinct spikes 
observed at $\hjd^\prime \sim 8666.1$ and 8672.8. In addition to these spikes, weak bumps are 
present just before the first spike and after the second spike, adding further complexity to 
the structure of the light curve.

Through 2L1S modeling, we identified a unique solution that precisely describes the observed 
features of the light curve. The binary parameters associated with this solution are $(s, q) 
\sim (1.32, 0.40)$, and the event has a time scale of $\te \sim 8.4$ days. This makes it the 
shortest-duration event among the reported anomalous microlensing events. The complete set of 
lensing parameters derived from this model is presented in Table~\ref{table:seven}. Additionally, 
the model curve corresponding to this solution is superimposed on the observed data points in 
Figure~\ref{fig:twentyfour}.

As shown in the lens system configuration, presented in the inset of the lower panel, the 
binary lens forms a resonant caustic that extends along the binary lens axis. The source 
entered the caustic by crossing the upper-right fold and exited through the lower-left fold. 
The two weak bumps observed before and after the caustic spikes occurred as the source passed 
through regions of positive deviation formed near the on-axis cusps of the caustic. The 
normalized source radius was precisely determined from the resolved caustics.

\begin{figure}[t]
\includegraphics[width=\columnwidth]{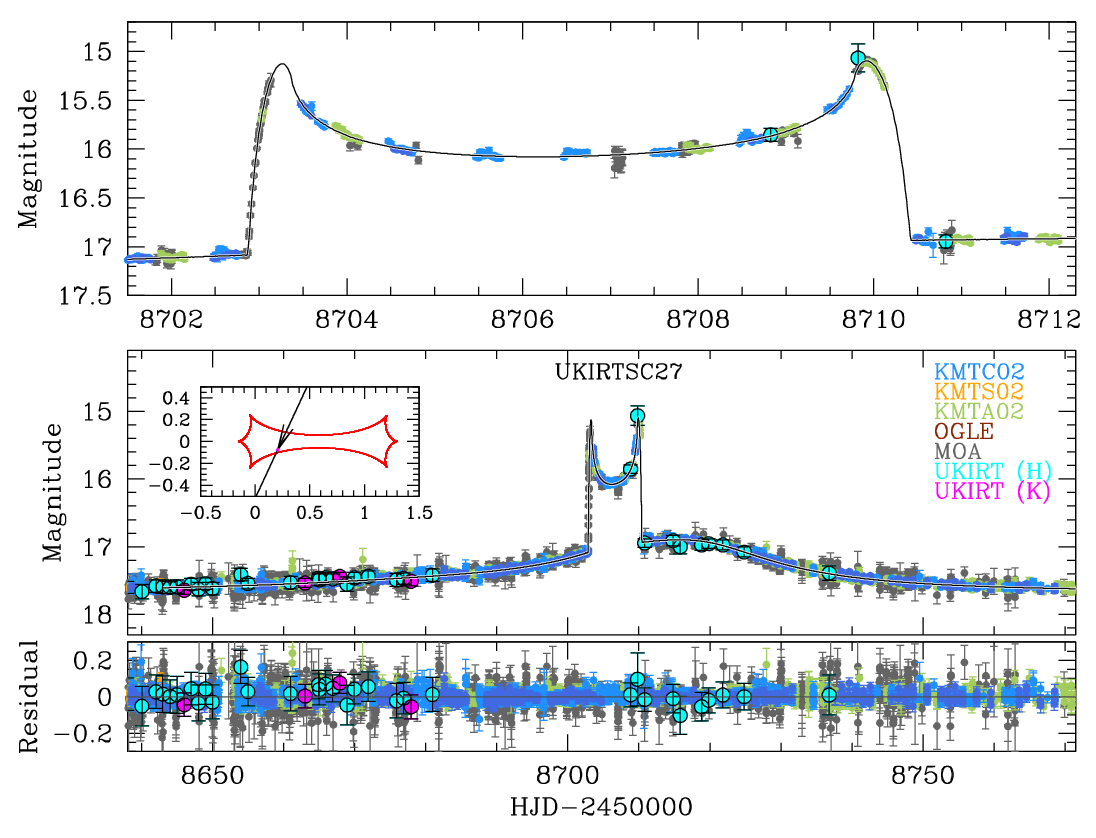}
\caption{
The lensing light curve of UKIRT27. 
}
\label{fig:twentyfive}
\end{figure}

\begin{table*}[t]
\caption{Angular source radius, Einstein radius, and relative proper motion.  \label{table:eight}}
\begin{tabular}{llllll}
\hline\hline
\multicolumn{1}{c}{Event}                   &
\multicolumn{1}{c}{$(V-I, I)_0$ }           &
\multicolumn{1}{c}{$\theta_*$ ($\mu$as) }   &
\multicolumn{1}{c}{$\thetae$ (mas) }        &
\multicolumn{1}{c}{$\mu$ (mas/yr)}      \\
\hline
 UKIRT02 & $(0.892 \pm 0.065, 18.252 \pm 0.020)$  &  $0.862 \pm 0.082 $  &  $0.890 \pm 0.086 $  &  $15.89 \pm 1.53 $ \\
 UKIRT04 & $(0.753 \pm 0.082, 18.427 \pm 0.013)$  &  $0.682 \pm 0.073 $  &  $0.131 \pm 0.019 $  &  $1.662 \pm 0.251$ \\
 UKIRT05 & $(1.061 \pm 0.134, 18.661 \pm 0.020)$  &  $0.86 \pm 0.13   $  &  $1.50 \pm 0.23   $  &  $5.84 \pm 0.88  $ \\
 UKIRT14 & $(0.931 \pm 0.043, 16.350 \pm 0.021)$  &  $2.17 \pm 0.18   $  &  $0.73 \pm 0.06   $  &  $7.43 \pm 0.61  $ \\
 UKIRT26 & $(1.491 \pm 0.041, 15.038 \pm 0.020)$  &  $6.24 \pm 0.51$  &     $0.387 \pm 0.032 $  &  $16.85 \pm 1.37 $ \\
 UKIRT27 & $(1.194 \pm 0.045, 15.721 \pm 0.020)$  &  $3.83 \pm 0.32   $  &  $0.529 \pm 0.044 $  &  $5.75 \pm 0.48  $ \\
\hline
\end{tabular}
\end{table*}

\subsection{UKIRT27} \label{sec:four-twentyfour}

The UKIRT lensing survey concluded with the completion of the 2019 season, and UKIRT27 marks 
the final anomalous event reported from the survey.  The source of the event lies in a region 
also monitored by other optical lensing surveys, including OGLE, MOA, and KMTNet. 
Figure~\ref{fig:twentyfive} presents the lensing light curve of the event, which features 
prominent caustic-crossing spikes at  $\hjd^\prime \sim 8703.0$ and 8710.2. The first caustic 
spike was resolved by the combined data from the MOA and KMTA datasets, while the second spike 
was covered by the KMTC, KMTA, and UKIRT $H$-band datasets.  In addition to the caustic spikes, 
the light curve displays a weak bump following the second caustic crossing, before gradually 
returning to the baseline.

The anomalies in the lensing light curve are well described by a 2L1S model. The lensing parameters 
for this solution are provided in Table~\ref{table:seven}. The binary lens is defined by parameters 
$(s, q) \sim (2.73, 1.01)$, producing a resonant caustic that extends parallel to the binary lens 
axis. The event has an estimated time scale of $\te \sim 33$ days.

The lens system configuration is illustrated in the inset of the lower panel of 
Figure~\ref{fig:twentyfive}. The source entered the caustic through the upper fold and exited 
via the lower fold, generating the distinct caustic spikes observed in the light curve. After 
the second spike, a weak bump was detected, caused by the source passing through a region of 
positive deviation near the lower-left cusp of the caustic. The normalized source radius was 
precisely measured by analyzing the resolved caustic crossings.

\begin{figure}[t]
\includegraphics[width=\columnwidth]{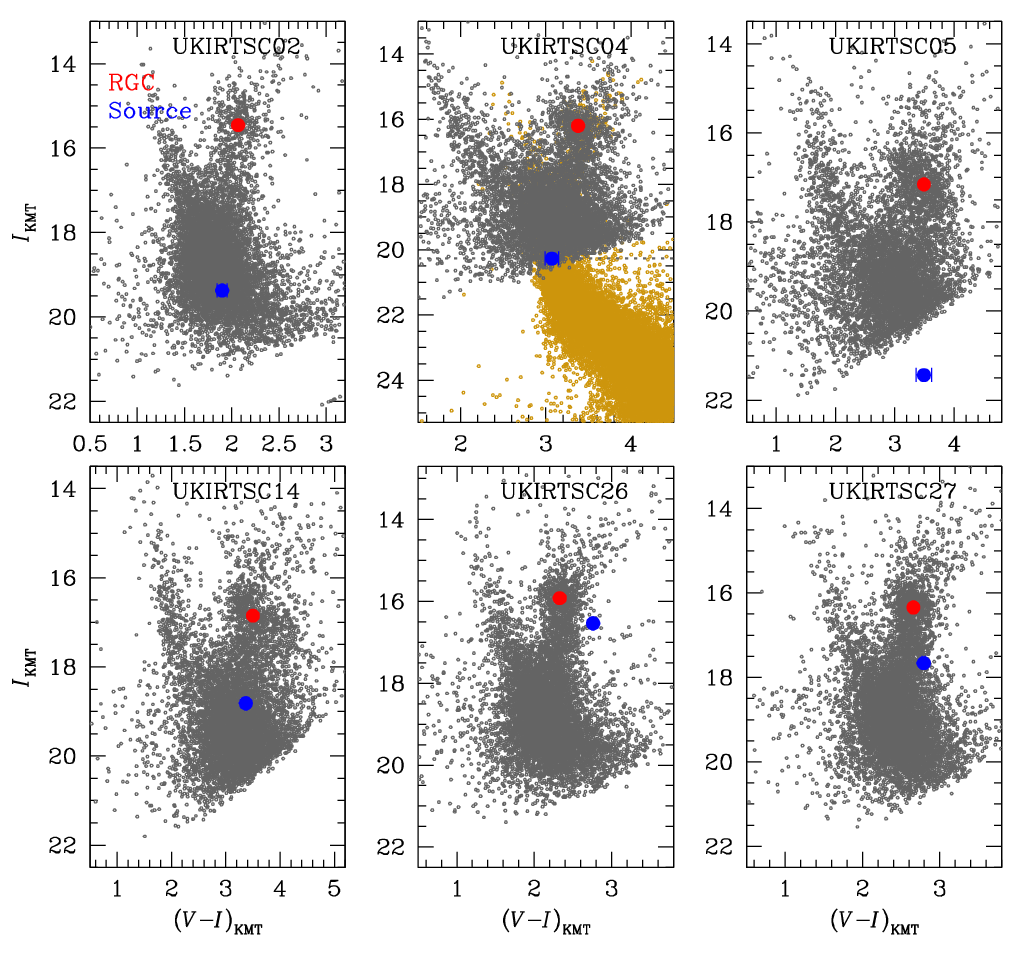}
\caption{
Locations of source stars for the lensing events UKIRT02, UKIRT04, UKIRT05, UKIRT14, UKIRT26, 
and UKIRT27.  The red filled dot in each panel represents the centroid of the red giant clump 
(RGC), which was used for calibrating the color and magnitude.
}
\label{fig:twentysix}
\end{figure}

\section{Angular Einstein radii} \label{sec:five}

In this section, we determine the Einstein radii for events with well-resolved caustics 
and well-covered $I$- and $V$-band light curves. The process begins by measuring the 
instrumental (uncalibrated) $I$- and $V$-band magnitudes of the source by fitting the 
light curve data to the model curve.  The light curve for measuring the source magnitude 
is constructed using the pydia software developed by \citet{Albrow2017}.  The source is 
then placed on the instrumental (uncalibrated) color-magnitude diagram (CMD) of nearby 
stars, which is constructed using the same pydia code.  To account for extinction and 
reddening, the source color and magnitude are calibrated using the centroid of the red 
giant clump (RGC) as a reference. For the dereddened color, $(V-I)_{{\rm RGC},0}$, and 
magnitude, $I_{{\rm RGC},0}$, of the RGC centroid, we adopt the values from \citet{Bensby2013} 
and \citet{Nataf2013}, respectively. The calibrated $V-I$ color is then converted into a 
$V-K$ color using the color-color relation from \citet{Bessell1988}. The angular radius 
of the source ($\theta_*$) is subsequently derived from the $(V-K)$–$\theta_*$ relation 
provided by \citet{Kervella2004}. Finally, the Einstein radius is calculated using the 
angular source radius and the normalized source radius through the relation $\thetae = 
\theta_*/\rho$. The relative lens-source proper motion is then determined as $\mu = 
\thetae/\te$, using the measured Einstein radius and the event time scale.

The Einstein radius was determined for six events: UKIRT02, UKIRT04, UKIRT05, UKIRT14, 
UKIRT26, and UKIRT27. For the remaining events, determining $\thetae$ proved challenging 
due to several factors. Firstly, many events observed exclusively by the UKIRT survey 
had data in only a single passband, and even when observations were made in two bands, 
the light curve coverage in one of them was often inadequate for accurate color 
determination. Secondly, for some events with measured $H-K$ color, calibration was 
difficult due to the lack of prior dereddened values for the RGC, hindering its use 
as a reference. Thirdly, in most UKIRT-only events, the light curve did not display a 
caustic crossing, and when one occurred, it was not sufficiently resolved to determine 
$\rho$. Finally, even for events where the caustic was resolved through additional survey 
data, the quality of $V$-band data was often inadequate for color determination due to 
significant extinction in regions near the Galactic center.

Figure~\ref{fig:twentysix} shows the locations of source stars for the six lensing events 
with measured Einstein radii. In the case of UKIRT04, for which the $I$-band source magnitude 
was determined but the $V$-band magnitude was not, we first aligned the CMD from ground-based 
observations with that obtained from HST observations \citep{Holtzman1998}. We then inferred 
the source color and magnitude as the mean values along the main-sequence branch. In 
Table~\ref{table:eight}, we listed the estimated values of the dereddened source color and
 magnitude, $(V-I, I)_0$, angular source radius, angular Einstein radius, and relative 
lens-source proper motion.

For the event UKIRT05, both the angular Einstein radius and the microlens parallax were 
measured. In this case, the physical parameters of the lens mass and the lens distance 
can be uniquely determined using the \citet{Gould2000} relation:
\begin{equation}
M = {\thetae \over \kappa\pie };\qquad
\dl = {{\rm AU} \over \pie\thetae + \pi_{\rm S}}.
\label{eq2}
\end{equation}
Here $\kappa = 4G/(c^2{\rm AU})$, $\pi_{\rm S} = {\rm AU}/\ds$ represents the parallax of 
the source lying at a distance $\ds$.  The estimated masses of the lens components and 
distance are
\begin{equation}
\eqalign{
M_1 & = (1.21 \pm 0.21)~M_{\odot}, \cr
M_2 & = (0.40 \pm 0.07)~M_{\odot}, \cr
\dl & = (3.37 \pm 0.40)~{\rm kpc}.
}
\label{eq3}
\end{equation}

\section{Summary and conclusion} \label{sec:six}

We conducted an investigation into the anomalous microlensing events reported by the UKIRT 
microlensing survey, which was carried out over a period of four years, from 2016 to 2019. 
This survey primarily focused on observing stars near the Galactic center using near-infrared 
passbands, with the goal of examining how microlensing event rates vary spatially across the 
field of the Galactic center. In the analysis, we incorporated data from 11 microlensing 
events that were also observed by other optical microlensing surveys, including those conducted 
by the OGLE, KMTNet, and MOA collaborations.

Among the 27 anomalous events reported from the survey, we revealed the nature of 24 events 
except for three events, in which UKIRT03 was likely to be a transient variable, and UKIRT10 
and UKIRT24 were were difficult to accurately characterize their nature due to the limitations 
of the available data. We have confirmed the binary lens nature of the anomalies for 21 events, 
including UKIRT02, UKIRT04, UKIRT05, UKIRT06, UKIRT07, UKIRT08, UKIRT09, UKIRT11, UKIRT12, 
UKIRT13, UKIRT14, UKIRT15, UKIRT16, UKIRT18, UKIRT19, UKIRT20, UKIRT22, UKIRT23, UKIRT25, 
UKIRT26, and UKIRT27. Among these binary lens events, the companion of UKIRT11L is a planetary 
object which has a mass ratio to its host of $(1.88 \pm 0.18) \times 10^{-3}$. For two events 
UKIRT01 and UKIRT17, the anomalies could be interpreted using either a binary-source or a 
binary-lens model.  For the precise description of the anomaly in UKIRT21, it was needed a model 
where not only the lens and but also source are binaries.

For the events UKIRT05 and UKIRT14, it was found that accounting for higher-order effects 
induced by the orbital motions of both Earth and the binary lens was crucial. While accurately 
determining the higher-order parameters for UKIRT14 proved challenging due to the difficulty 
in distinguishing between microlens-parallax and lens-orbital effects, these parameters were 
securely measured for UKIRT05. With these measurements, along with the angular Einstein radius, 
the component masses of the UKIRT05 binary lens were determined to be $M_1 = (1.05 \pm 
0.20)~M_\odot$, $M_2 = (0.36 \pm 0.07)~M_\odot$, and the distance to the lens was found to be 
$\dl = (3.11 \pm 0.40)$~kpc.

The UKIRT lensing experiment was primarily conducted to support the preparation of future 
lensing surveys with the Roman telescope. This study demonstrated that lensing events 
can be successfully detected even in the central regions of the Galaxy, where optical 
observations are challenging. However, the observational conditions, such as precision 
and cadence, differ between the UKIRT and Roman experiments. Moreover, the limited number 
of detected events in the UKIRT data makes it insufficient for statistically estimating 
the frequency of higher-order effects or the probability of events requiring complex 
modeling, such as 2L2S or 3L1S.

\begin{acknowledgements}
This research was supported by the Korea Astronomy and Space Science Institute under the 
R\&D program (Project No. 2025-1-830-05) supervised by the Ministry of Science and ICT.
This research has made use of the KMTNet system operated by the Korea Astronomy and Space Science 
Institute (KASI) at three host sites of CTIO in Chile, SAAO in South Africa, and SSO in Australia. 
Data transfer from the host site to KASI was supported by the Korea Research Environment Open NETwork 
(KREONET).  This research was supported by the KASI under the 
R\&D program (Project No. 2023-1-832-03) supervised by the Ministry of Science and ICT.
The MOA project is supported by JSPS KAKENHI Grant Number
JP24253004, JP26247023,JP16H06287 and JP22H00153.
J.C.Y., I.G.S., and S.J.C. acknowledge support from NSF Grant No. AST-2108414. 
C.R. was supported by the Research fellowship of the Alexander von Humboldt Foundation.
W. Zang and H.Y.  acknowledge support by the National Natural Science Foundation of China (Grant 
No. 12133005). W.Zang acknowledges the support from the Harvard-Smithsonian Center for Astrophysics 
through the CfA Fellowship.
\end{acknowledgements}

\end{document}